\documentclass[aps,prd,10pt,twocolumn,superscriptaddress,floatfix,nofootinbib,amsmath,amssymb,altaffilletter,preprintnumbers]{revtex4-1}
\usepackage[colorlinks=true,pdfstartview=FitV,breaklinks=true]{hyperref}

\usepackage{bm}
\usepackage{graphicx}
\usepackage{multirow}
\usepackage{soul}
\usepackage{amsmath}
\usepackage{fontawesome}
\usepackage{mathrsfs}
\usepackage{amssymb}
\usepackage{yfonts}
\usepackage{color}
\usepackage{xspace}
\usepackage{url}
\usepackage{verbatim}
\usepackage{mathtools}
\usepackage{upgreek}
\usepackage{units}
\usepackage{siunitx}
\usepackage{amstext}
\usepackage[sc]{mathpazo}
\usepackage{booktabs}
\usepackage{tabulary}
\usepackage{tabularx}
\usepackage{etoolbox}
\usepackage[version=4]{mhchem}
\usepackage[utf8]{inputenc}
\usepackage[dvipsnames,table]{xcolor}
\newcolumntype{Y}{>{\centering\arraybackslash}X}
\hypersetup{urlcolor=BlueViolet,
	    citecolor=Plum,
	    linkcolor=PineGreen}
\usepackage[official]{eurosym}
\definecolor{lightgray}{rgb}{0.9,0.9,0.9}	    
\definecolor{green}{rgb}{0,0.5,0}
\definecolor{red}{rgb}{1,0,0}
\definecolor{blue}{rgb}{0,0,0.5}

\newcommand{\dbd}[2]{\ifmmode \frac{\textrm{d}#1}{\textrm{d}#2}\else $\textrm{d}#1/\textrm{d}#2$\fi}
\newcommand{\pbp}[2]{\ifmmode \frac{\partial#1}{\partial#2}\else $\partial#1/\partial#2$\fi}

\newcommand{\ra}[1]{\renewcommand{\arraystretch}{#1}}

\DeclareMathAlphabet{\mathpzc}{OT1}{pzc}{m}{it}
 \newcommand{\eV}{\text{e\kern-0.15ex V}\xspace}

 \newcommand{\TeV}{\text{T\kern-0.1ex \eV}\xspace}

\DeclareMathAlphabet{\mathpzc}{OT1}{pzc}{m}{it}

\newcommand{\gag}{g_{a\gamma}}

\newcommand{\be}{\begin{equation}}
\newcommand{\ee}{\end{equation}}
\newcommand{\bea}{\begin{eqnarray}}
\newcommand{\eea}{\end{eqnarray}}

\def\nn{\nonumber}

\begin{document}

\title{Axion helioscopes as solar magnetometers}

\author{Ciaran A. J. O'Hare}\email{ciaran.ohare@sydney.edu.au}
\affiliation{School of Physics, Physics Road, The University of Sydney, NSW 2006 Camperdown, Sydney, Australia}
\author{Andrea Caputo}\email{andrea.caputo@uv.es}\affiliation{Instituto de Fisica Corpuscular, Universidad de Valencia and CSIC, Edificio Institutos Investigacion, Catedratico Jose Beltran 2, Paterna, 46980 Spain}
\author{Alexander J. Millar}\email{alexander.millar@fysik.su.se}\affiliation{The Oskar Klein Centre for Cosmoparticle Physics,
Department of Physics,
Stockholm University, AlbaNova, 10691 Stockholm, Sweden}
\affiliation{Nordita, KTH Royal Institute of Technology and
Stockholm
  University, Roslagstullsbacken 23, 10691 Stockholm, Sweden}
\author{Edoardo Vitagliano}\email{edoardo@physics.ucla.edu}\affiliation{ Department  of  Physics  and  Astronomy,  University  of  California,  Los  Angeles,  California,  90095-1547,  USA}

\preprint{NORDITA-2020-055}

\date{\today}
\smallskip
\begin{abstract}
Axion helioscopes search for solar axions and axion-like particles via inverse Primakoff conversion in strong laboratory magnets pointed at the Sun. Anticipating the detection of solar axions, we determine the potential for the planned next-generation helioscope, the International Axion Observatory (IAXO), to measure or constrain the solar magnetic field. To do this we consider a previously neglected component of the solar axion flux at sub-keV energies arising from the conversion of longitudinal plasmons. This flux is sensitively dependent to the magnetic field profile of the Sun, with lower energies corresponding to axions converting into photons at larger solar radii. If the detector technology eventually installed in IAXO has an energy resolution better than 200 eV, then solar axions could become an even more powerful messenger than neutrinos of the magnetic field in the core of the Sun. For energy resolutions better than 10 eV, IAXO could access the inner 70\% of the Sun and begin to constrain the field at the tachocline: the boundary between the radiative and convective zones.
The longitudinal plasmon flux from a toroidal magnetic field also has an additional 2\% geometric modulation effect which could be used to measure the angular dependence of the magnetic field. \smallskip \href{https://github.com/cajohare/solax}{\large\faGithub}
\end{abstract}

\maketitle

\section{Introduction} 
\label{sec:intro}


Axions~\cite{Peccei:1977hh, Peccei:1977ur, Weinberg:1977ma, Wilczek:1977pj, Kim:2008hd} and axion-like particles~\cite{Jaeckel:2010ni} (hereafter referred to as axions) are light pseudoscalars with extremely weak couplings to Standard Model fields. Although by construction axions are difficult to observe, they are expected to possess an experimentally advantageous coupling to the photon, $\gag$. In the absence of any accidental cancellations, this coupling is a generic feature of the most well-known and well-motivated models known as `QCD axions'~\cite{DiLuzio:2020wdo}, which are involved in the solution of Peccei and Quinn to the strong CP problem~\cite{Peccei:1977hh, Peccei:1977ur}. This coupling is the most widely-studied as it facilitates axion-photon conversion inside magnetic fields~\cite{Sikivie:1983ip, Raffelt:1987im, vanBibber:1988ge}.

Experimental searches for axions via their photon coupling fall broadly into three categories: {\it light-shining-through-wall} experiments, which search for axions converting to and from photons either side of an opaque barrier~\cite{Ballou:2015cka, DellaValle:2015xxa, Bahre:2013ywa,Ehret:2010mh,Betz:2013dza}; {\it haloscopes}, which search for the cold population of axions that could constitute the dark matter halo of the Milky Way~\cite{Hagmann:1990tj,DePanfilis:1987dk,Asztalos:2010,Brubaker:2016ktl,Ouellet:2018beu,TheMADMAXWorkingGroup:2016hpc,Alesini:2019ajt,Alesini:2017ifp,Goryachev:2017wpw,McAllister:2017lkb,McAllister:2018ndu,Du:2018uak,Braine:2019fqb,Boutan:2018uoc,Lee:2020cfj,Zhong:2018rsr,Gramolin:2020ict}; and {\it helioscopes}, the subject of this work, which search for the thermal flux of axions expected to be emitted by the Sun~\cite{Semertzidis:1990qc,Moriyama:1998kd,Inoue:2008zp,Zioutas:1998cc,Zioutas:2004hi,Andriamonje:2007ew,Arik:2008mq,Arik:2011rx,Arik:2013nya,Anastassopoulos:2017ftl}. See Ref.~\cite{Irastorza:2018dyq} for a recent overview of experimental searches.

A helioscope consists of a long magnet bore pointed at the Sun, with the field aligned in the transverse direction. When the flux of $\sim$keV solar axions passes through this applied field, a small number will convert into UV--x-ray photons. The expected number of signal photon events is given by a convolution of the solar axion flux with the axion-photon conversion probability which is maximized for long, strong, and uniform magnetic fields. The CERN Axion Solar Telescope (CAST)~\cite{Zioutas:1998cc,Zioutas:2004hi,Andriamonje:2007ew,Arik:2008mq,Arik:2011rx,Arik:2013nya,Anastassopoulos:2017ftl} ran from 2003--2017 with a field of up to 9.5~T and a length of 9.26~m. It set a final constraint of $\gag>6.6\times 10^{-11}$~GeV$^{-1}$~\cite{Anastassopoulos:2017ftl}, the strongest limit to date for the majority of the axion mass range below $m_a\sim10$~meV. The next generation helioscope, the International Axion Observatory (IAXO)~\cite{Irastorza:2013dav, Armengaud:2014gea}, aims to have sensitivity to couplings more than an order of magnitude below CAST: well into the territory of QCD axion models between $m_a\sim 10^{-3}$--$1$~eV. 

There are substantial prospects for IAXO to discover the QCD axion or one of the multiplicity of proposed axion-like particles~\cite{Masso:1995tw,Masso:2002ip,Ringwald:2012hr,Ringwald:2012cu,Arvanitaki:2009fg,Cicoli:2012sz,Jaeckel:2010ni}, as well as to resolve numerous model dependence issues relating to said models if the axion is detected first elsewhere (see e.g.~Ref.~\cite{Armengaud:2019uso}). Here we instead consider the potential for IAXO to perform highly novel post-discovery \emph{solar} physics by constraining the poorly understood magnetic field of the Sun. The prospects were brought to light recently in Ref.~\cite{Caputo:2020quz} which calculated a previously ignored component of the solar axion flux due to the conversion of longitudinal plasmons. 

The flux of axions from longitudinal plasmons is distinct from other sources of solar axions --- such as the Primakoff, and ``ABC'' fluxes --- and has several interesting differences. The flux is arises as a resonance at the plasma frequency at a particular solar radius. This means that the frequency of the axions generated by the conversion of longitudinal plasmons can be directly mapped to the particular solar position at which they converted. Additionally, and most importantly for this study, the size of the flux from a particular position is proportional to the square of the magnetic field at that position. Taken together we can see that the longitudinal plasmon flux provides an almost direct way to measure the \emph{profile} of the solar magnetic field, where previously only bounds over certain regions of the Sun were possible. 

In this paper we show that longitudinal plasmon flux could be extremely useful probe of the solar magnetic field profile, as well as just a new source of axions. Indeed, this work follows several previous studies of the potential for post-discovery physics and astrophysics with axions~\cite{Jaeckel:2019xpa,OHare:2017yze,Knirck:2018knd,OHare:2018trr,OHare:2019qxc}. In the event of discovery\footnote{Recently an excess of low energy electronic recoils was reported by the XENON1T collaboration~\cite{XENON1T} with a spectrum consistent with solar axions. Although we caution that the axion couplings required to fit the signal are ruled out by several astrophysical bounds on stellar cooling~\cite{AxionElectron_GC,Viaux:2013lha,Hansen_RG} and SN1987A~\cite{Carenza:2019pxu}}, axions  --- together with neutrinos --- could establish a powerful multimessenger astroparticle probe of the invisible interior of our Sun. In

The structure of the paper is as follows: in Sec.~\ref{sec:solaraxions} we review the calculation of the two principal fluxes of solar axion relevant for the coupling $\gag$: Primakoff and longitudinal plasmon conversion. Then in Sec.~\ref{sec:solarBfield} we review the status of our understanding of the solar magnetic field so that we can place a window on potential models. In Sec.~\ref{sec:heliscopes} we overview the formalism for calculating the signal in a helioscope and summarize the experimental parameters envisaged for IAXO. We present our results in Sec.~\ref{sec:results} and our statistical approach in Appendix~\ref{sec:stats}.

The results presented in this work are reproducible via publicly available python notebooks\footnote{\url{https://github.com/cajohare/solax}}.

\section{Solar axions}\label{sec:solaraxions}

If an axion exists then the Sun may be a powerful factory for them, generating them in its interior through several different processes. Via the coupling to the photon, $g_{a\gamma}$, the most widely considered process is Primakoff conversion: photons converting into axions in the electromagnetic fields of the electrons and ions making up the solar plasma. This flux is dominant in hadronic QCD axion models like the KSVZ~\cite{Kim:1979if,Shifman:1979if}, but is a major component of the flux for DFSZ models as well~\cite{Dine:1981rt,Zhitnitsky:1980tq}.\footnote{Non-hadronic models also possess a tree-level coupling to electrons, so in these cases the axion flux has a large contribution from ``ABC'' processes: atomic recombination and deexcitation, bremsstrahlung, and Compton scattering~\cite{Redondo:2013wwa}. For this work we will only be interested in $\gag$ so do not consider this flux.} Here we extend the set of solar axion fluxes to include the flux from the conversion of longitudinal plasmons.

To calculate any flux of solar axions one must assume a solar model. Following previous work on this~\cite{Caputo:2020quz}, and related subjects~\cite{Redondo:2013wwa,Vitagliano:2017odj}, we use the Saclay model~\cite{Couvidat:2002bs,TurckChieze:2001ye} which provides\footnote{At \href{http://irfu.cea.fr/dap/Phocea/Vie_des_labos/Ast/ast_visu.php?id_ast=1444}{this http url}.} the temperature $T(r)$, and electronic and ionic densities, $n_e(r)$ and $n_i(r)$, as a function of radial position, $r$, in the solar interior.

\subsection{Primakoff flux}
\begin{figure}[t]
\includegraphics[width=0.49\textwidth]{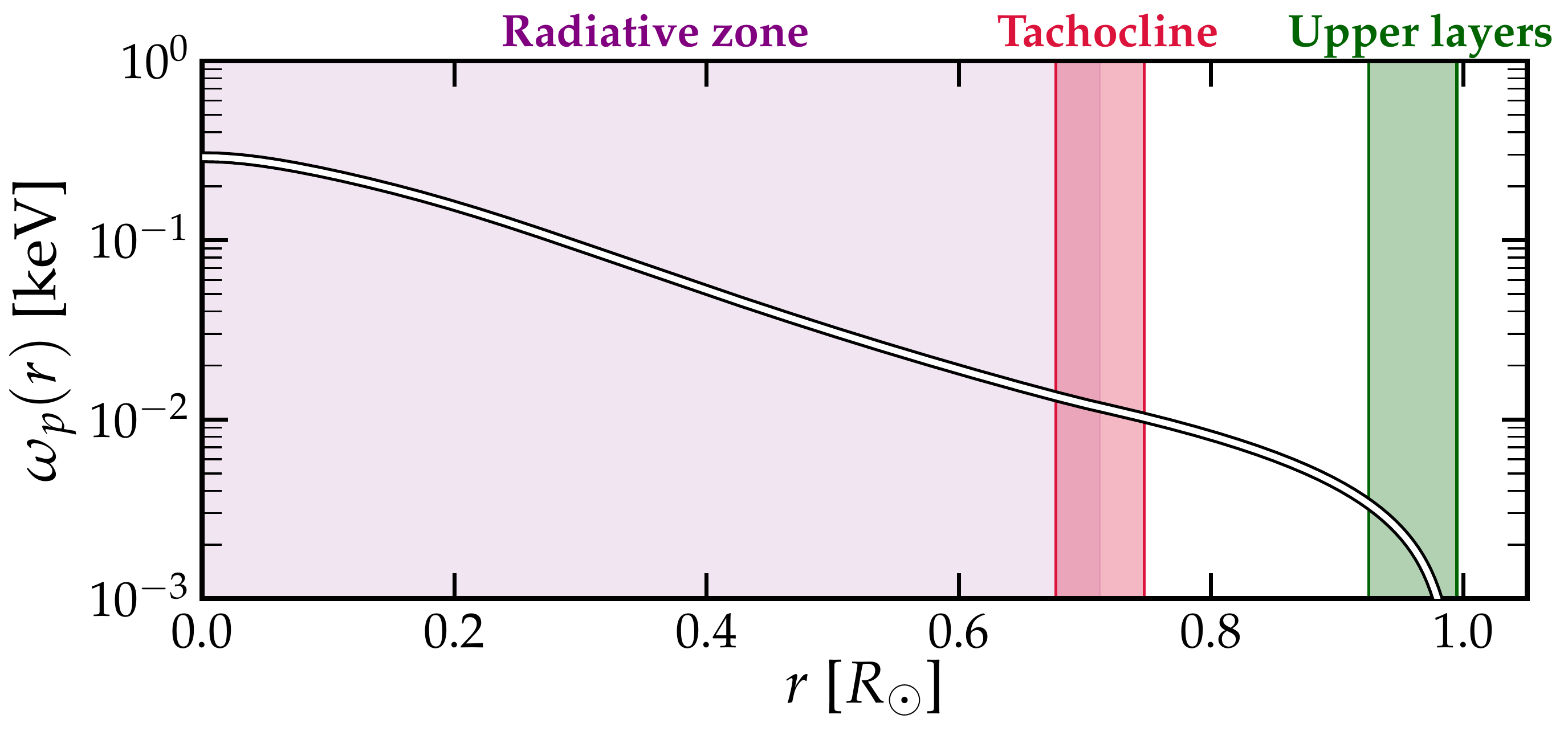}
\caption{\label{fig:SaclayData_wp} The solar plasma frequency $\omega_p = \sqrt{4\pi \alpha n_e/m_e}$ as a function of solar radius, assuming the Saclay model~\cite{Couvidat:2002bs,TurckChieze:2001ye}. We also label in purple, red and green respectively three regions of the seismic $B$-field model in which the magnetic field is important: the inner radiative zone, the tachocline, and the upper layers. These will be important when we discuss the longitudinal plasmon flux which will originate from these particular ranges of radii.}
\end{figure}

For the axion flux from Primakoff processes we will not use the commonly adopted empirical formula, but instead make a slightly refined calculation, as in Refs.~\cite{Caputo:2020quz,Jaeckel:2006xm}, which accounts for the nonzero plasma frequency, $\omega_p(r)$. This only affects the flux for axion energies below $\omega \lesssim 300$~eV so is a small correction to the final signal --- but is a necessary refinement when also considering the flux from longitudinal plasmon conversion also present at these energies. The rate of Primakoff conversion to axions with frequency $\omega$ is written as~\cite{Jaeckel:2006xm},
\begin{equation}
\Gamma_{\gamma \rightarrow a}(\omega,r)=\frac{g_{a \gamma}^{2} k_{s}^{2} T}{64 \pi} \int_{-1}^{1} \textrm{d} \cos \theta \frac{\sin ^{2} \theta}{(x-\cos \theta)(y-\cos \theta)} \, ,
\end{equation}
where,
\begin{align}
    x(r)&=\left(k_{a}^{2}+k_{\gamma}^{2}(r)\right)/2 k_a k_\gamma(r) \, , \\ \nn
    y(r)&=x(r)+k_{s}^{2}(r) / 2 k_{a} k_{\gamma}(r) \, .
\end{align}
For ultrarelativistic axions we can assume the dispersion relation $k_a \approx \omega$. The screening scale $k_s(r)$ is given by the Debye-H\"uckel formula which sums over electrons ($e$) and ions ($i$) with atomic numbers $Z_i$,
\begin{equation}
k_{\mathrm{s}}^{2}(r)=\frac{4 \pi \alpha}{T(r)}\left(n_{e}(r)+\sum_{i} n_{i}(r) Z_{i}^{2} \right)\, .
\end{equation}
The fine structure constant is given by $\alpha$. The dispersion relation for the transverse plasmons involved in this process is written: $k_\gamma(r)= \omega^2 -\omega^2_p(r)$, where $\omega_p(r)$ is the plasma frequency,
\begin{equation}
    \omega_p^2=\frac{4\pi\alpha n_e}{m_e} \, .
\end{equation}
This imposes a restriction on the range of possible axion energies generated by Primakoff processes: $\omega>\omega_p(r)$. Therefore the refinement of allowing $\omega_p \neq 0$ only leads to a suppression of the flux relative to the simplified calculation. The plasma frequency as a function of radius is shown in Fig.~\ref{fig:SaclayData_wp}.

We then integrate the conversion rate $\Gamma_{\gamma \rightarrow a}(\omega,r)$ over the Sun to calculate the flux of axions arriving at Earth with energy $\omega$,
\begin{equation}\label{eq:Primakoff_flux}
\frac{\mathrm{d} \Phi_P}{\textrm{d} \omega}=\frac{1}{(1\,{\rm AU})^2} \int_{0}^{R_{\odot}} r^{2} \textrm{d} r \frac{\omega^{2}}{\pi^{2}} \frac{\Gamma_{\gamma\rightarrow a}(\omega,r)}{\mathrm{e}^{\omega / T(r)}-1} \, .
\end{equation}

\begin{figure}[t]
\includegraphics[width=0.49\textwidth]{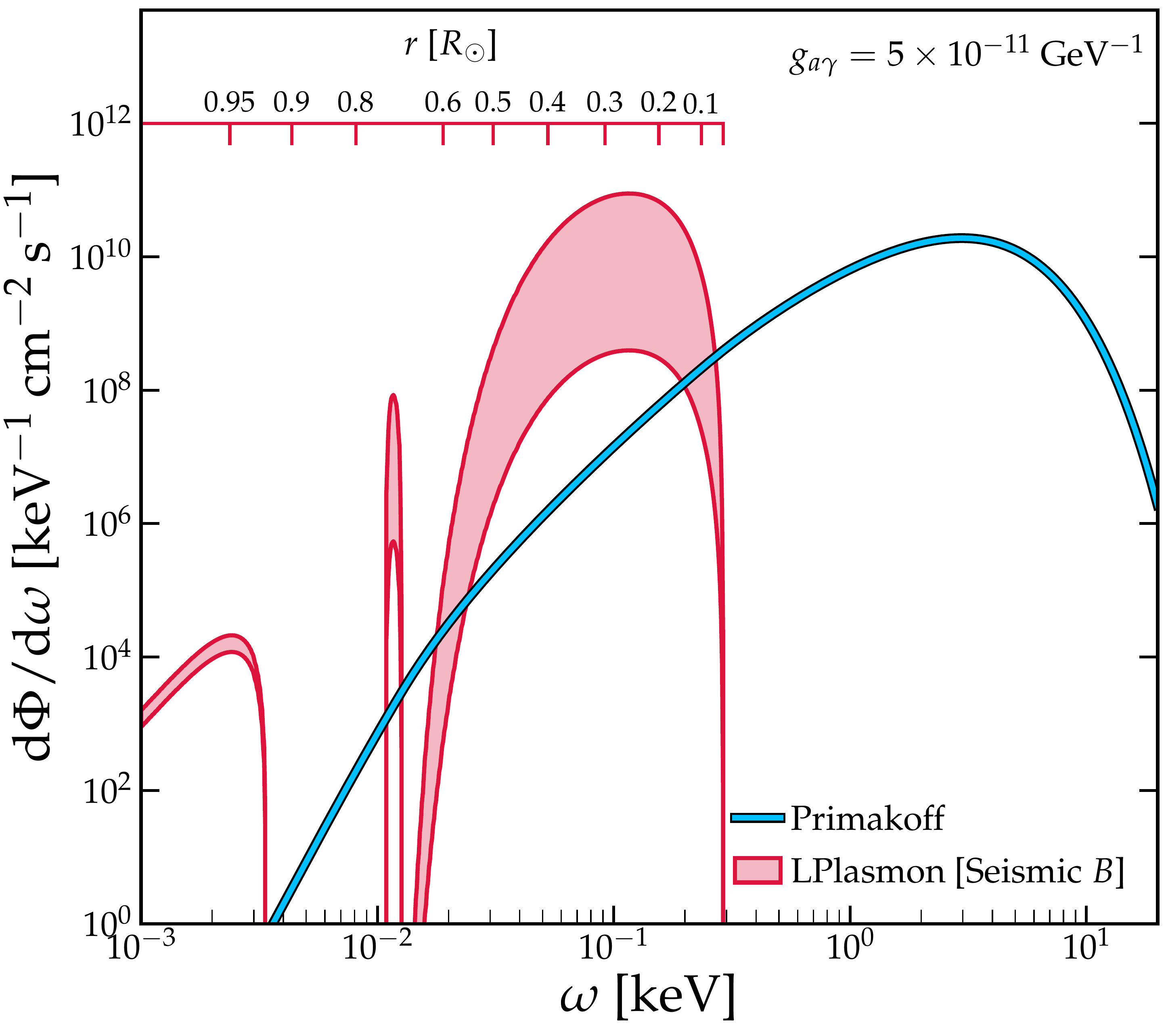}
\caption{\label{fig:PlasmonFluxes}
Flux of axions arriving at Earth originating from the two process we study here: the Primakoff effect (blue) and the conversion of longitudinal plasmons (red). For both fluxes we use the plasma frequency profile $\omega_p(r)$ calculated under the Saclay standard solar model, and for the LPlasmon flux we assume the seismic solar magnetic field profile from Ref.~\cite{Couvidat:2003ba} but with varying normalizations. The solar radius from which the LPlasmon flux originates can be obtained by inverting $\omega = \omega_p(r)$, we show this scale with an auxiliary horizontal axis in red.}
\end{figure}
This flux is shown as a blue line on Fig.~\ref{fig:PlasmonFluxes} assuming $\gag = 5\times 10^{-11}$~GeV$^{-1}$. It is dominant for $\omega\gtrsim~\omega_p(r~=~0)~\approx~0.3$~keV.

\subsection{Plasmon conversion flux}
The next source of axions, and the one which is our primary concern here, is the flux from the conversion of longitudinal plasmons (LPlasmons) sourced by the solar magnetic field $B(r)$. The conversion of LPlasmons to new particles in the presence of strong magnetic fields has been considered in the past for a variety of models and environments, see e.g.~Refs.~\cite{Mikheev:1998bg,Hardy:2016kme,Redondo:2013lna,Pospelov:2008jk}. The calculation was most recently revised for the LPlasmon-axion conversion in the Sun in Ref.~\cite{Caputo:2020quz} using thermal field theory. We summarize quickly the calculation here before discussing the solar magnetic field, which is the primary source of uncertainty --- and what we would like to determine with IAXO.

The rate of LPlasmon-axion conversion is obtained by relating the axion self-energy in the presence of a magnetic field (proportional to $\gag^2$) to the corresponding photon damping rate of transverse and longitudinal quanta $\Gamma_{\rm T,L}$~\cite{Raffelt:1996wa, Redondo:2013wwa}. From the pioneering work of Weldon~\cite{Weldon:1983jn} we know in general that the emission rate of a boson by a thermal medium is related to the self-energy of the particle in the medium, $\Pi$, via~\cite{Weldon:1983jn, Kapusta:2006pm},
\begin{equation}\label{eq:emissionrate_selfenergy}
	\Gamma= -\frac{{\rm Im} \, \Pi}{\omega(e^{\omega/T}-1)}.
\end{equation}
The self-energy we are interested in here is the longitudinal part of the axion's: ${\rm Im} \,\Pi_{a,{\rm L}}$. This is dependent on the component of the magnetic field parallel to the axion momentum, ${B_{\|}=\hat{\mathbf{k}} \cdot \mathbf{B}}$, and reads:
\begin{equation}
	{\rm Im}\, \Pi_{a,{\rm L}} = m_a^2 g^2_{a\gamma} B^2_{\|}\,{\rm Im}\left(\frac{Z_L}{\omega^2-\omega^2_p -i Z_L {\rm Im}\, \Pi_{\gamma,L}}\right),
\end{equation}
where we have introduced the vertex renormalization constant ${Z_L = \omega^2 / (\omega^2-\textbf{k}^2)}$ and the equilibrium LPlasmon self-energy ${{\rm Im}\, \Pi_{\gamma,L} = - Z_L^{-1} \omega \Gamma_L}$. The four-momentum of the external axion is packaged as $(\omega,\textbf{k})$. Plugging these into the emission rate formula Eq.\eqref{eq:emissionrate_selfenergy}, we find the rate of axions produced from LPlasmons~\cite{Caputo:2020quz,Mikheev:1998bg}:
\begin{equation}
\Gamma_{\text{LP}\rightarrow a}(\omega)=\frac{g_{a \gamma}^{2} B_{\|}^{2}}{e^{\omega / T}-1} \frac{\omega^{2} \Gamma_{L}}{\left(\omega^{2}-\omega_{p}^{2}\right)^{2}+\left(\omega \Gamma_{L}\right)^{2}} \, .
\end{equation}
This formula exhibits a resonance at the plasma frequency $\omega_p$. Approximating the resonance with a delta function, $\delta(\omega-\omega_p)$\footnote{We have checked that the out-of-resonance production of axions is negligible. This is expected in the narrow-width approximation for the plasmon in which the real part of the polarization function is much larger than the imaginary part.}, we can then integrate over phase space and over the Sun to get the luminosity in axions,
\begin{equation}\label{eq:luminosity}
L_{\text{LP}\rightarrow a} = \int_{\odot} \mathrm{d}^{3} \mathbf{r} \int \frac{\mathrm{d}^{3} \mathbf{k}}{(2 \pi)^{3}} \omega \frac{g_{a \gamma}^{2} B_{\|}^{2}}{(e^{\omega / T}-1)} \frac{\pi}{2} \delta\left(\omega-\omega_{p}\right) \, .
\end{equation}

If, for now, we assume spherical symmetry for the magnetic field $B_{\|}(r) = B(r)$, the flux at Earth at a distance of $r_\oplus \approx 1$ AU away from the Sun is obtained as,
\begin{align}\label{eq:LPflux_spherical_symmetry}
\frac{\mathrm{d} \Phi_{\mathrm{LP}}}{\mathrm{d} \omega}&=\frac{1}{4 \pi r^2_\oplus} \int_{\odot} \mathrm{d}^{3} \mathbf{r} \frac{\omega^{2}}{(2 \pi)^{3}} \frac{g_{a \gamma}^{2} B^{2}(r)}{e^{\omega / T(r)}-1} \frac{2 \pi^{2}}{3}\, \delta\left(\omega-\omega_{p}(r)\right) \nonumber \\
&=\frac{1}{12 \pi r^2_\oplus} \int_{0}^{R_{\odot}} \mathrm{d}r \, r^{2} \frac{\omega^{2} g_{a \gamma}^{2} B(r)^{2}}{e^{\omega / T(r)}-1} \delta\left(\omega-\omega_{p}(r)\right) \nonumber \\
&=\frac{1}{12 \pi r^2_\oplus}\, r_{0}^{2}\, \frac{\omega^{2} g_{a \gamma}^{2} B^2\left(r_{0}\right)^{}}{e^{\omega / T\left(r_{0}\right)}-1} \frac{1}{\left|\omega_{p}^{\prime}\left(r_{0}\right)\right|} \, .
\end{align}
where in the final step $r_0$ is given by inverting ${\omega = \omega_p(r_0)}$, and ${\omega^\prime_p(r) = \textrm{d}\omega_p(r)/\textrm{d} r}$.

Thus we see that --- thanks to the resonance --- axions with a particular energy originate from a particular solar shell. This fact is precisely why we anticipate that axions could be a valuable probe of the solar magnetic field. In effect, a measurement of the axion flux at to decreasing energies can be recast as a measurement of the solar magnetic field at increasing radii. We discuss the details of the solar magnetic field in the next section but for now we show an example of the LPlasmon-axion flux in Fig.~\ref{fig:PlasmonFluxes} (red) along with the Primakoff flux (blue). For this particular example of a motivated solar magnetic field profile, we can see that for certain energy ranges below 0.3~keV the LPlasmon flux dominates, even for the most conservative field normalizations. For context, in this range we also indicate parallel to frequency, the value of the radius from which the LPlasmon originates, i.e.~the value that satisfies $\omega = \omega_p(r)$ following the line on Fig.~\ref{fig:SaclayData_wp}.

\section{Solar magnetic field}\label{sec:solarBfield}
\subsection{Solar models}
All the necessary ingredients to compute the solar axion flux are given in a standard solar model (SSM), a simplified description of the Sun~\cite{kippenhahn1990stellar,Serenelli:2016dgz,Vitagliano:2019yzm}.
To build an SSM, one usually assumes spherical symmetry  and  hydrostatic equilibrium. Any dynamical effects, rotation, or magnetic  fields are neglected. Energy  is  transported  by  radiation (i.e.~photons), and in the outer 30\% (by radius) by convection as well. The former is described using theoretical opacity calculations like OPAL~\cite{Iglesias:1996bh} or OP~\cite{Badnell:2004rz}, whereas the latter is treated phenomenologically through a  mixing  length parameter. The Sun is evolved from a homogeneous zero-age model defined by a set of adjustable parameters which are chosen to match the present-day age, luminosity, size, and surface metal to hydrogen ratio. 

All of the solar model properties of the Sun are known with great accuracy, with the exception of its chemical composition. There are a variety of empirical methods of determining the Sun's chemistry, from the infrared-optical spectrum of the photosphere, the x-ray spectrum of the corona, as well as the abundances of CI chondritic meteorites~\cite{Bergemann:2014vaa}. Furthermore, with helioseismology --- the study of acoustic sound wave propagation inside the Sun --- we can determine the sound-speed profile, the depth of the convective zone, as well as the surface helium abundance.

In the past, theoretical determinations of solar abundances, like the well-known ``GS98'' calculation~\cite{Grevesse:1998bj} found surprisingly good agreement with helioseismology despite many simplifying assumptions. More modern composition calculations on the other hand, for example the ``AGSS09'', while more sophisticated in construction, ultimately found abundances that were in substantial tension with observed data~\cite{Asplund:2009fu,Scott:2014lka,Scott:2014mka,Grevesse:2014nka}. This issue has not yet been resolved, and is known as ``solar abundance problem''~\cite{Serenelli:2009yc}. The current status of the problem is summarized in Ref.~\cite{Villante:2019tcd}.

Nevertheless, the  temperature  and  plasma frequency  profiles  of  the  Sun  predicted  by  different  solar  models  are consistent to a good degree.  We therefore choose to use the reference seismic Saclay  model~\cite{TurckChieze:2001ye, Couvidat:2003ba},\footnote{Sometimes, the expression ``seismic model'' is also used for models in which one takes the sound speed-profile 
and the density 
as an input, computes the pressure 
assuming hydrostatic equilibrium, and finally derives the temperature 
and the helium abundance 
with some additional input physics. On the other hand, the model built in Ref.~\cite{Couvidat:2003ba} is based on standard 1D stellar evolution codes to compute a model in agreement with seismic observation.} as  it  is  the most complete publicly available model concerning the external layers. Uncertainty in the plasma frequency and the temperature stemming from the SSM abundance problem is less than a few percent~\cite{Serenelli:2009yc}. So for fluxes dependent on these parameters, this propagates to an uncertainty of less than 10\%~\cite{Vitagliano:2017odj}. Therefore we can safely focus on the solar magnetic field uncertainty, which will almost certainly be larger.


\subsection{The magnetic field}
While SSMs are quasi-static simplifications of a star, they can be augmented to include dynamical effects and information about the magnetic field. In fact this can be important as magnetic pressure will affect the sound-speed, the density as well as neutrino fluxes~\cite{Couvidat:2003ba}. 

The magnetic field of the Sun is most important in three separate regions: the central radiative zone ($r\lesssim 0.7 R_\odot$); the boundary between the radiative and convective zones known as the tachocline ($r\sim 0.7 R_\odot$); and the upper layers ($r\gtrsim 0.9$).

The most easily detectable LPlasmon flux will be from the central radiative zone, in which the central magnetic field may be extremely high. The topology of this field is likely to be mostly toroidal, however it has been shown that a purely toroidal field would be subject to instabilities~\cite{tayler1973adiabatic,braithwaite2006stability}. A poloidal component would act to stabilize the field~\cite{tayler1980adiabatic,braithwaite2009axisymmetric,Duez:2009vg}, however if this component were too large it would penetrate the convective zone and cause a large unobserved asymmetry between the two halves of the magnetic cycle. This argument implies a strong bound of $\lesssim 10^2$--$10^3$ G on the poloidal part of the radiative zone's magnetic field, but not on the toroidal part. Both toroidal and poloidal components would have originated from a small seed poloidal field present in the pre-main sequence phase of the Sun's evolution~\cite{Friedland:2002is}.

\begin{figure*}[t]
\includegraphics[width=\textwidth]{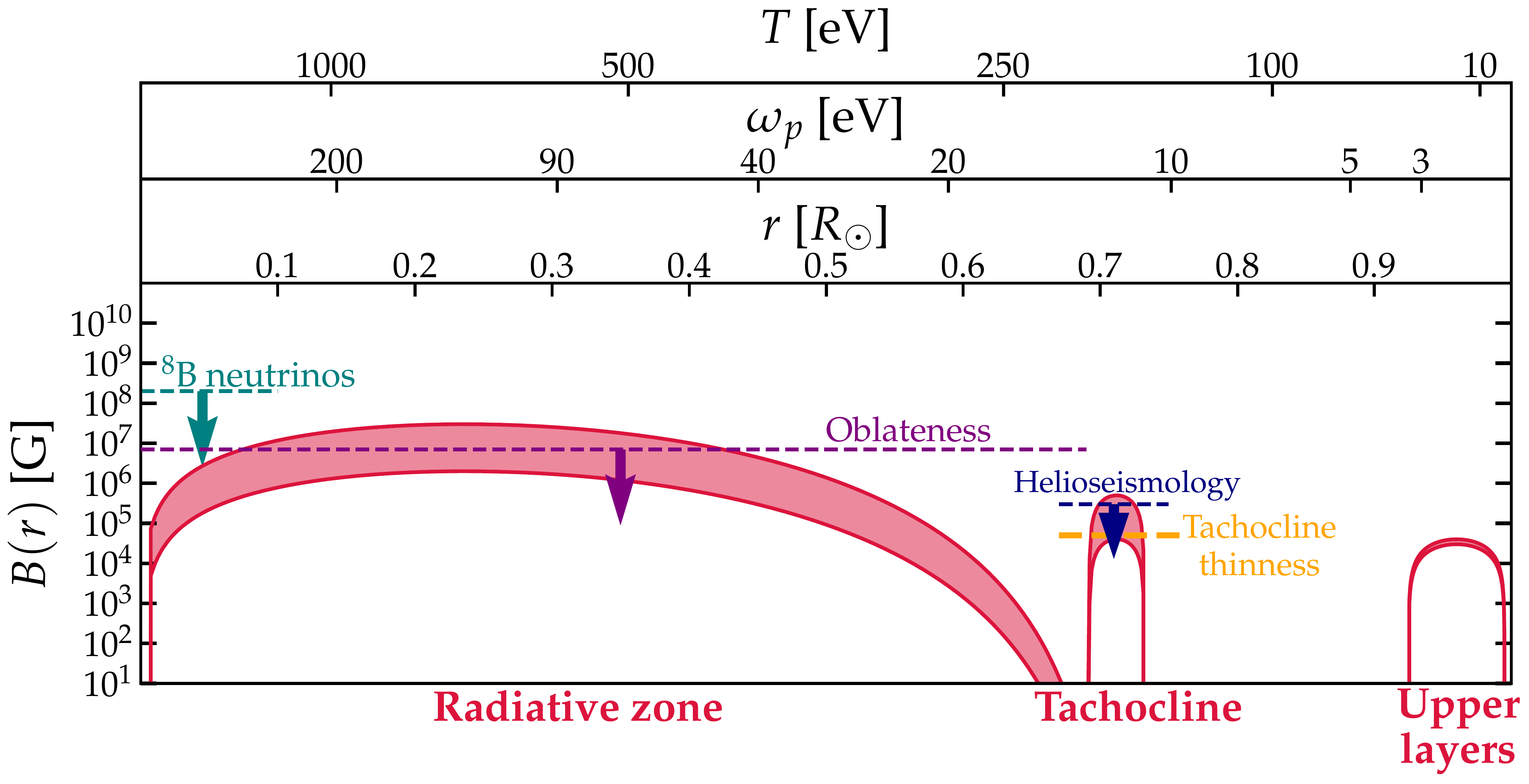}
\caption{\label{fig:Bfield} Solar magnetic field $B(r)$ as a function of radius $(r)$, plasma frequency $(\omega_p)$ or temperature, $(T)$. The three distinct regions of the seismic solar $B$-field model are labelled. For each we bound the possible magnetic field strengths between our most and least conservative values discussed in Sec.~\ref{sec:solarBfield}. These windows are: $B_{\rm rad} \in [2,30] \times 10^6$ G for the radiative zone; $B_{\rm tach} \in [4,50]\times 10^4$ G for the tachocline; and $B_{\rm upper}~\in~[3,4]\times~10^4$~G for the upper layers. We also show as horizontal lines some of the constraints on certain parts of the magnetic field discussed in Sec.~\ref{sec:solarBfield}.}
\end{figure*}

In the Saclay model, the magnetic field is assumed to be purely toroidal~\cite{Couvidat:2003ba},
\begin{align}\label{eq:Bfield_angular}
    \mathbf{B}(\mathbf{r})=B(r)\frac{\textrm{d}}{\textrm{d}\theta}P_k(\cos{\theta}) \, \hat{\mathbf{e}}_\phi
\end{align}
in heliocentric spherical coordinates $(r,\theta,\phi)$, where $P_k(\cos{\theta})$ are Legendre polynomials and $\hat{\mathbf{e}}_\phi$ a unit vector pointing in the azimuthal direction. We assume a quadrupolar field, i.e.~$k=2$, in accordance with surface-magnetism observations~\cite{Couvidat:2002bs, Baldner_2009}. In Ref.~\cite{Caputo:2020quz} the Legendre polynomial was neglected, reducing the flux by a factor of 1.2.  The radial profile is given by,
\begin{widetext}
\begin{equation}
 B(r) =
  \begin{cases}
    B_{\rm rad}(1+\lambda)(1+\frac{1}{\lambda})^{\lambda}\left(\frac{r}{r_{\rm rad}}\right)^{2}\left[1-\left(\frac{r}{r_{\rm rad}}\right)^{2}\right]^{\lambda} & \text{for } r<r_{\rm rad} \ ,\ {\lambda\equiv 10\, r_{\rm rad}+1} \\
    B_{\rm tach}\left[1-\left(\frac{r-r_{\rm tach}}{d_{\rm tach}}\right)^{2}\right] & \text{for } |r-r_{\rm tach}|<d_{\rm tach} \\
    B_{\rm outer}\left[1-\left(\frac{r-r_{\rm upper}}{d_{\rm upper}}\right)^{2}\right] & \text{for } |r-r_{\rm upper}|<d_{\rm upper}\\
	0& \text{for } r>R_\odot \, .
  \end{cases} \quad 
\end{equation}
\end{widetext}

The three regions describe the radiative, tachocline and outer regions of the Sun. For the radiative zone we use $r_{\rm rad} = 0.712 R_\odot$; the tachocline is centered on $r_{\rm tach}=0.712 R_{\odot}$ and has half-width thickness $d_{\rm tach}=0.035 R_{\odot}$; finally, the upper layers are centered on ${r_{\rm upper}=0.96 R_{\odot}}$ with $d_{\rm upper}~=~0.035 R_{\odot}$. These three regions can be identified from left to right in Fig.~\ref{fig:Bfield}. For context we also indicate the scales of $\omega_p$ and $T$ for the corresponding solar radius, $r$. Since the LPlasmon flux as a function of energy traces the radial profile of the magnetic field via $\omega_p(r)$, the three distinct regions are also visible in the flux, as was shown in Fig.~\ref{fig:PlasmonFluxes}.

The normalizations of the magnetic fields $(B_{\rm rad},B_{\rm tach},B_{\rm upper})$ present the greatest uncertainty. One can consider, for example, the benchmark values of Ref.~~\cite{Couvidat:2003ba} in which the authors vary the magnetic field strengths over the ranges: $B_{\rm rad}= 1$--$5 \times 10^7$~G; $B_{\rm tach} =3$--$5\times10^5$~G and $B_{\rm upper} = 2$--$3\times 10^4$~G. However, other studies indicate that this range for $B_{\rm rad}$ is too high. 
For example, Ref.~\cite{Friedland:2002is} claims a tighter bound on the magnetic field in the radiative zone using measurements of the solar oblateness:
\begin{equation}
	B_{\rm{rad}} \lesssim 7\times 10^6 ~\rm G \, .
\end{equation}

The tachocline is the name given to the transition layer between the radiative and convective zones. In the radiative zone the rotation rate is uniform whereas the convective zone has differential rotation. The range of field strengths of the tachocline $3$--$5\times 10^5$~G is in reasonable agreement with helioseismology~\cite{Antia2000,2002ApJ...578L.157C,Baldner_2008,Baldner_2009,Kiefer_2018}. These studies are a mixture of both upper bounds~\cite{Antia2000,Baldner_2009} and fits which are improved by the presence of a magnetic field~\cite{Baldner_2008,2002ApJ...578L.157C}. 

A different approach is to simulate the structural changes to the Sun associated with a magnetic field~\cite{Barnab__2017}. Structurally, the tachocline ought to be a wide boundary because of radiative spreading, also known as differential rotation burrowing~\cite{1991JAtS...48..651H, SpiegelZahn}. Surprisingly though the tachocline is observed to be rather narrow. A potential explanation for this found by Ref.~\cite{Barnab__2017} is if the magnetic field's Lorentz force acts to suppress the differential rotation. However this explanation invokes a slightly lower value for the field than the aforementioned values, at $B_{\rm tach}\sim 5 \times 10^4$ G. So we will use this value instead for our lower bound in the tachocline.

Finally, for the upper layers we merely remark here that this flux is likely to be unobservable for the foreseeable future. Not only would this require energy resolutions at the $\sim 1$~eV level, the overall flux itself is extremely weak. We will present results on this region for completeness, but do not consider field strengths larger than the $B_{\rm outer}= 4\times 10^4$ G of the seismic model.

So to summarize, we will bound a window of reasonable field normalizations.  Based on the discussion above, we choose
\begin{align}\label{eq:Bfield_window}
    B_{\rm rad} \in [2,\,30]\times 10^6 \, {\rm G} \, , \nonumber \\
    B_{\rm tach} \in [4,\,50]\times 10^4 \, {\rm G} \, ,  \\
    B_{\rm upper} \in [3,\,4]\times 10^4 \, {\rm G} \, .\nonumber
\end{align}
In Fig.~\ref{fig:Bfield} the solar magnetic field profile for the seismic model is shown bounded over these ranges. We emphasize that while the magnetic field of the Sun is certainly nonzero, the lower bounds we have chosen here are only partially based on existing constraints, and are otherwise somewhat arbitrary. Our motivation for setting a benchmark window --- rather than simply an upper bound --- is to allow us to demonstrate in a clearer fashion how the projected $B$-field sensitivity of IAXO is dependent on the size of the field normalization. In that spirit we fix this window of values for all results presented here.

This rather large window of possible values demonstrates why an axionic measurement of the magnetic field could be so valuable. The LPlasmon flux is an almost direct probe of the field's \emph{profile}, whereas other probes have to date only been able to put bounds on the field in certain radial ranges.

\subsection{Complementary measurements}
All the measurements and inferences of the solar magnetic field discussed so far suffer greatly from uncertainties, and are sometimes drawn under different assumptions. Therefore a direct measurement, or even just a complementary constraint on the strengths of these fields would be highly sought after.  
We predict that the sensitivity of IAXO may easily overcome any other probes of the strength of $B$.

In particular, around the core of the Sun ${r \lesssim 0.09\, r_{\odot}}$, where helioseismology loses accuracy, one can consider bounds from the $\ce{^{8}B}$ neutrino flux. The presence of a magnetic field modifies the pressure inside the star with an additional $p_{\rm magnetic}=B^2/8\pi$; this implies a change in the gas temperature $\delta T/T \sim p_{\rm magnetic}/p_{\rm gas}$. The neutrino flux from $\ce{^{8}B}$ decay is very sensitive to the core temperature (as the flux scales as $T_c^{24\pm5}$~\cite{Bahcall:1996vj}), which leads to the upper limit~\cite{Friedland:2002is}
\begin{equation}
	B_{\rm core} \lesssim 2\times 10^8 ~\rm G \, .
\end{equation}
Since this field is considerably stronger than the value we assumed in examples presented so far for the radiative zone normalization, we can already see that solar axions could provide a much stronger bound than neutrinos. 

Possibly the most novel aspect of the solar axion LPlasmon flux is its direct dependence on the radial profile of the field, something which is not immediately accessible with other techniques. Moreover, the solar axion flux at higher energies corresponds to smaller radii, so measurements of the field in the core of the Sun are in fact the least demanding of a detector's energy resolution. In summary, the LPlasmon flux should be an ideal tool to measure the magnetic field in the central regions of the Sun and would be highly complementary, as we will demonstrate quantitatively in Sec.~\ref{sec:results}.

\subsection{Angle and time dependence}
In Eq.\eqref{eq:LPflux_spherical_symmetry} we assumed spherical symmetry for the solar magnetic field. Given that a more realistic topology for the field is likely to be toroidal, we should check the effect this has on the LPlasmon flux.

To determine this, we go back to Eq.\eqref{eq:luminosity} and leave incomplete both the angular integral over the magnetic field and the wave vector of the emitted axion and the volume integral over the Sun. Instead we can write down the rate of axion emission from a particular point in the Sun
\begin{align}\label{eq:Naxions_LPlasmon_angularintegral}
	\frac{\textrm{d}N_{\rm ax}}{\textrm{d}t \, \textrm{d}\omega\textrm{d}V}&=\int\frac{\textrm{d}\Omega_k}{(2\pi)^3} \frac{ g_{a\gamma}^2 B_{\|}^{2}\omega^2}{e^{\omega/T}-1} \frac{\pi}{2}\delta(\omega-\omega_p(r))  \\ &=\frac{ g_{a\gamma}^2B(r,\theta)^2\omega^2\delta(\omega-\omega_p(r))}{16\pi^2 (e^{\omega/T(r)}-1)} \int \textrm{d}\Omega_k (\hat{\bf {k}}\cdot{\hat{\bf B}})^2 \,. \nonumber
\end{align}
The remaining delta function will be used to integrate over the volume of the Sun.
We now assume that the magnetic field has a toroidal configuration, like Eq.\eqref{eq:Bfield_angular} which was in the form: $\mathbf{B} = B(r,\theta) \hat{\mathbf{e}}(\phi)$. Adopting heliocentric Cartesian coordinates and using spherical angles $(\theta,\phi)$ for the magnetic field and $(\theta_k,\phi_k)$ for the axion, we can write, 
\begin{align}
    \hat{\bf B} &= \hat{\mathbf{e}}_\phi = (-\sin\phi,\,\cos\phi,\,0) \, , \nonumber \\
    \hat{\bf{k}} &= (\sin\theta_k \cos\phi_k,\,\sin\theta_k \sin\phi_k,\,\cos\theta_k ) \,, \nonumber \\
    &\Rightarrow \hat{\bf {k}}\cdot{\hat{\bf B}} = \sin\theta_k\sin(\phi_k-\phi) \, .
\end{align}
As we are interested in the flux passing through IAXO, we can treat the Sun as pointlike and only worry about a very small range of angles, $\theta_\oplus<\theta_k<\theta_\oplus+\delta$ where $\theta_\oplus$ is heliocentric polar angle position of the Earth. The small angle we need to integrate over is $\delta \approx d/r_\oplus$ where $d$ is some small test distance in the polar direction (for example, the size of IAXO). This defines a cylindrical surface for the flux to pass through of height $d$ and radius $r_\oplus\sin\theta_\oplus$. Integrating over the full azimuthal angle will allow us to simplify the problem via the circular symmetry of the magnetic field, which ensures that any point on the cylinder will experience the same flux.

The azimuthal part of the solid angle integral in Eq.(\ref{eq:Naxions_LPlasmon_angularintegral}) is simply, 
\begin{equation}
    \int \mathrm{d}\phi_k\,(\hat{\bf {k}}\cdot{\hat{\bf B}})^2 = \pi \sin^2\theta_k,
\end{equation}
which can then be integrated over the polar angle,
\begin{align}
    \int \textrm{d}\Omega_k (\mathbf{k}\cdot \hat{\mathbf{B}})^2 &=
    \pi\int_{\theta_\oplus}^{\theta_\oplus+\delta}  \mathrm{d}\theta_k \sin^3\theta_k \nonumber \\
    &\approx \pi\delta   \sin^3\theta_\oplus  \nonumber \\
    &= \frac{\pi d}{r_\oplus}   \sin^3\theta_\oplus \, .
\end{align}

Now we can write down the differential flux at Earth by considering the fact that the axions from a given $(r,\phi,\theta)$ will be emitted over a surface $S =~d~\times~2\pi r_\oplus\sin\theta_\oplus$: 
\begin{align} 
    \frac{\mathrm{d}\Phi_{LP}}{\mathrm{d}\omega \textrm{d}V}(r,\phi,\theta) &= \frac{1}{S} \frac{\textrm{d}N_{\rm ax}}{\textrm{d}t \, \textrm{d}\omega \, \textrm{d}V} \\ &= \frac{\sin^2\theta_\oplus}{32 \pi^2 r^2_\oplus} \frac{ g_{a\gamma}^2B^2(r,\theta)\omega^2}{(e^{\omega/T(r)}-1)}\delta(\omega-\omega_p(r)), \nonumber
\end{align}
We then finally integrate over the solar volume to get the final flux we will use,
\begin{equation}
    \frac{\mathrm{d}\Phi_{LP}}{\mathrm{d}\omega} = \int_\odot \textrm{d}^3\mathbf{r} \frac{\mathrm{d}\Phi_{LP}}{\mathrm{d}\omega \textrm{d}V}(r,\phi,\theta) \, .
\end{equation}
Compared to the spherically symmetric assumption Eq.\eqref{eq:LPflux_spherical_symmetry}, the toroidal configuration gains us a factor of $3\sin^2\theta_\oplus/2$ due to the preferential emission towards the Earth which orbits approximately above the solar equator $\sin\theta_\oplus \approx 1$.

\begin{figure*}[t]
\includegraphics[width=\textwidth]{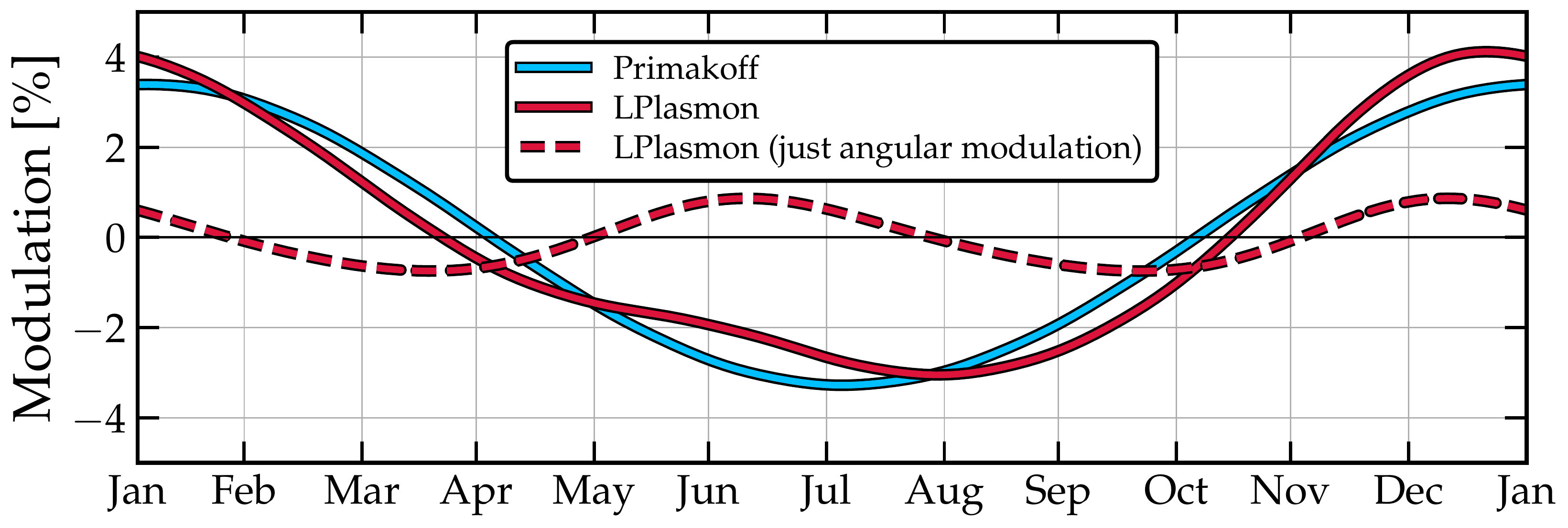}
\caption{\label{fig:Modulation} Percentage modulation in the Primakoff and LPlasmon signals over one year. The modulation of the Primakoff signal is due solely to the changing Earth-Sun distance, whereas the LPlasmon flux modulates with distance \emph{and} the position of the Earth with respect to the solar equator. Because of the toroidal shape of the magnetic field, axions from LPlasmon conversion are preferentially emitted from the equator. The LPlasmon flux therefore modulates biannually and is a maximum when the ecliptic plane crosses the solar equatorial plane --- the modulation due to this effect alone is shown as a dashed line.}
\end{figure*}

But while $\sin\theta_E\approx 1$ is approximately true, the Sun's rotation axis is in fact slightly tilted from the ecliptic by $7.25^\circ$. Because of this we expect that the LPlasmon flux from a toroidal field will modulate by $\sim 1.6\%$ biannually. We show this modulation in Fig.~\ref{fig:Modulation}. We have calculated $\theta_\oplus(t)$ by first taking the right ascension and declination of the ecliptic as a function of time, $(\alpha_{\rm ecl}(t),\delta_{\rm ecl}(t))$, and then computing the angle between this plane and the solar north pole: $(\alpha^{\rm NP}_\odot,\delta^{\rm NP}_\odot) = (286.13^\circ,63.87^\circ)$. The modulation caused by $\sin^2{\theta_\oplus(t)}$ is shown as a dashed line in Fig.~\ref{fig:Modulation}.

However another source of time dependence that will affect both Primakoff and LPlasmon fluxes is an additional $\sim 7$\% annual modulation due to the change in the distance between the Earth and the Sun because of the Earth's orbital eccentricity:
\begin{equation}\label{eq:solarnu}
  \frac{1}{r^2_\oplus(t)} = \frac{1}{(1\, {\rm AU})^2} \left[ 1 + 2 e \cos\left(\frac{2\pi(t- t_e)}{T}\right) \right]
\end{equation}
where $e = 0.016722$ and $T = 1$ year and  $t_e = 3$ days relative to Jan 1. This source of time dependence is known with effectively negligible uncertainty so could be safely accounted for in some modulation analysis. Such an analysis may be of great interest since this time dependence would allow the experiment to access angular information about the magnetic field. For instance a measurement of the size of the modulation with respect to the total average LPlasmon flux would constrain the ratio between the toroidal and poloidal components of the field. If the magnetic field was highly poloidal for example, almost all the axions would be emitted away from the equatorial plane, leading to a large suppression of the signal. Since the modulation of a toroidal field is only a percent level effect we expect to need probably around 100 times more events from the LPlasmon flux than would be needed to initially measure the flux, so would only be possible for relatively large values of $\gag$. Nevertheless it could be of great interest post-discovery and we leave a detailed exploration of this to future work.

Before proceeding we reiterate here for clarity that our computation of the LPlasmon flux now differs from Ref.~\cite{Caputo:2020quz} and accounts for the Legendre polynomial form for the polar angle dependence of $B(r,\theta)$ as well as the preferential equatorial emission of axions. This leaves us with an additional factor of,
\begin{equation}\label{eq:angular_enhancement}
\bigg\langle \bigg(\frac{\textrm{d}}{\textrm{d}\theta} P_2(\cos{\theta})\bigg)^2 \bigg\rangle_{\cos{\theta}} \times \frac{3}{2} \langle \sin^2{\theta_\oplus}\rangle_t \approx 1.8
\end{equation}
which we multiply Eq.~\eqref{eq:LPflux_spherical_symmetry} by. 

\section{IAXO}\label{sec:heliscopes}
\begin{table*}[t!]
\ra{1.3}
\begin{tabularx}{0.78\textwidth}{lll|YYY}
\hline\hline
 & & & {\bf babyIAXO} & {\bf IAXO} & {\bf IAXO+} \\
\hline
Magnetic field	& $B_{\rm lab}$\quad & [T] \quad & 2.5 & 2.5 & 3.5 \\
Magnet length	& $L$ \quad& [m] \quad & 20 & 20 & 22 \\
Number of bores	& $n_B$\quad & \quad & 2 & 8 & 8 \\
Total aperture area	& $S$ \quad& [cm$^2$] \quad & 0.77 & 2.26 & 3.9 \\
Measurement time	& $t$ \quad& [years] \quad & 1.5  & 3  & 5 \\
Telescope efficiency & $\varepsilon_{\rm T}$\quad & \quad & 0.7 & 0.8 & 0.8 \\
Detector efficiency & $\varepsilon_{\rm D}$\quad & \quad& 0.35 & 0.7 & 0.7 \\
Detector area & $A_{\rm det}$ \quad & [cm$^2$] \quad & $2\times0.3$ & $8\times0.15$ & $8\times0.15$ \\
Background & $\Phi_{\rm bg}$\quad &  [keV$^{-1}$ cm$^{-2}$ s$^{-1}$] & $10^{-7}$ & $10^{-8}$ & $10^{-9}$ \\
\hline\hline
\end{tabularx}
\caption{Experimental parameters for the three stages of IAXO: the prototype babyIAXO, the baseline IAXO and the upgraded IAXO+.}
\label{tab:IAXOparams}
\end{table*}

\subsection{Helioscope formalism}
The next step is to take the fluxes we have written down and use them to calculate the signal from the back-conversion of the solar axions into photons inside a laboratory magnetic field. The standard design of a helioscope consists of a long magnetic bore pointed at the Sun with a collecting photodetector at one end. Since the setup relies on two instances of axion-photon conversion, the signal is ultimately proportional to $g_{a\gamma}^4$ for the Primakoff and LPlasmon fluxes. The expected number of photons reaching a detector placed at the end of the helioscope of length $L$ is given by the integral,
\begin{equation}\label{eq:Ngamma}
 N_{\gamma} = S\, t\, \int  \mathrm{d}\omega \, \varepsilon_{\rm D}(\omega)\varepsilon_{\rm T}(\omega)\left(\frac{{\rm d} \Phi_{\rm P}}{{\rm d} \omega}+\frac{{\rm d} \Phi_{\rm LP}}{{\rm d} \omega}\right) \, P_{a \rightarrow \gamma}(\omega)  \, ,
\end{equation}
where $S$ is the total cross-sectional area of the helioscope, and $t$ is the measurement time (given by the total experimental duration multiplied by the experiment's duty cycle of around 50\%). We allow for two efficiency functions describing the detector ($\varepsilon_{\rm D}$) and the telescope itself ($\varepsilon_{\rm T}$). 

When the magnet bores are filled with a vacuum, the axion-photon conversion probability has the form,
\begin{equation}
P_{a \rightarrow \gamma}(\omega) = \left(  \frac{g_{a \gamma}B_{\rm lab}}{q} \right)^{2} \sin^{2}\left( \frac{qL}{2} \right) \, ,
\label{Pvac}
\end{equation}
where $B_{\rm lab}$ is the helioscope's applied magnetic field, and $q=m_{a}^{2}/2\omega$ is the axion-photon momentum transfer. The conversion probability is maximized when the axion's mass is negligible, or at least very small, and the interference patterns caused by the mismatch in the axion and photon dispersion relations take place over scales much longer than the magnet length. In the above equation this is when $qL < \pi$. For $L\sim 20$~m, this coherent conversion condition holds up until around $m_a\sim$~meV. For larger masses the signal suffers from rapid oscillatory features which ultimately suppress the total rate of photon production. Helioscopes operating in a vacuum mode therefore steeply lose sensitivity to large axion masses.

This is unfortunate since for CAST and IAXO these masses correspond to QCD models. So to rescue the experimental sensitivity at higher masses the coherence condition must be somehow recovered. This can be achieved by filling the bore with a buffer gas such as $^4$He or $^3$He. The new medium provides an effective photon mass $m_\gamma = \sqrt{4\pi \alpha n_e/m_e}$, which for an appropriate choice of electron density can be made equal to $m_a$, thereby matching the photon and axion dispersion relations. When this occurs, the axions and photons remain in phase along the magnet length and the signal is amplified again.

In the buffer gas regime, the conversion probability is instead~\cite{vanBibber:1988ge} --
\begin{align}\label{eq:Pbuffergas}
P_{a \rightarrow \gamma} =& \left(  \frac{g_{a \gamma}B}{2} \right)^{2}  \frac{1}{q^{2}+ \Gamma^{2}/4} \nonumber \\
&\times  \left( 1 + e^{-\Gamma L} - 2 \, e^{-\Gamma L/2} \cos(qL) \right) \, ,
\end{align} 
where the momentum transfer is now, $q~=~|m_{a}^{2}~-~m_{\gamma}^{2}|/2\omega$, and $\Gamma$ is the inverse absorption length for photons in the buffer gas. A search over a range of $m_a$ can then be performed by tuning the photon mass, which in practical terms means tuning buffer gas pressure, $p$. For our purposes it is sufficient to approximate the inverse absorption length and photon mass with an empirical formula found from a fit~\cite{GalanThesis} to x-ray mass-attenuation-coefficient data on $^4$He~\cite{Xraydatabase},
\begin{align}
    \Gamma & \approx 0.29 \frac{p(\textrm{mbar})}{\omega(\textrm{keV})^{3.1} \, T_{\rm gas}({\rm K})} \,  {\rm m}^{-1}\\
    m_\gamma & \approx  \sqrt{0.02 \frac{p(\textrm{mbar})}{T_{\rm gas}({\rm K})}} \, {\rm eV} \, .
\end{align}

\subsection{Experimental parameters}
The numerical values we adopt for the experimental parameters entering Eq.\eqref{eq:Ngamma} are summarized in Table~\ref{tab:IAXOparams}. For the baseline IAXO, we use the configurations anticipated in the conceptual design report~\cite{Irastorza:2013dav, Armengaud:2014gea} which assumes eight bores (total $S=2.26$ m$^2$) and a 6 year total data-taking time (including both vacuum and buffer gas runs). We also show the expected parameters for the prototype ``babyIAXO'' and the upgraded version named ``IAXO+''.

The collaboration is considering several detector technologies for the focal planes of the telescope, optimized for energies between $10$~keV down to well below $1$~keV.  Several promising proposals are under investigation such as the well-established micromegas, with $\sim 200$~eV resolutions~\cite{Krieger:2018nit,Garza:2016nty}, as well as metallic magnetic calorimeters which have demonstrated a full width at half maximum resolution of $\sim$2 eV~\cite{Kempf:2018}. We implement the effect of the energy resolution in our calculations by convolving Eq.\eqref{eq:Ngamma} with a Gaussian of width $E_{\rm res}$. The energy threshold is also assumed to be equal to $E_{\rm res}$. Following Refs.~\cite{Irastorza:2013dav, Armengaud:2014gea} we assume the telescope and detector efficiency functions are flat in energy, with $\varepsilon_{\rm T} = 0.8$ and $\varepsilon_{\rm D} = 0.7$ respectively. 

We impose a hard threshold cutoff at $E<E_{\rm res}$, since we cannot assume any energy sensitivity below this. However this is still a simplification relative to the true final sensitivity of IAXO's detectors. The efficiency function $\varepsilon_{\rm D}$ will certainly have a dependence on energy which suffers at energies close to $E_{\rm res}$ however this will vary with the precise technology. So we adopt this simplification, which was done for IAXO's projections, and remain naive with regards to the final detector. This consequently means that our results must be interpreted not as a definitive prediction for the sensitivity of IAXO, but instead a guide towards the energy sensitivity required to probe the $B$-field to a given radius. In fact the study we present here provides substantial additional motivation for detectors with excellent sub-10~eV energy resolutions.

\subsection{Signal spectra}

\begin{figure*}[t]
\includegraphics[width=0.49\textwidth] {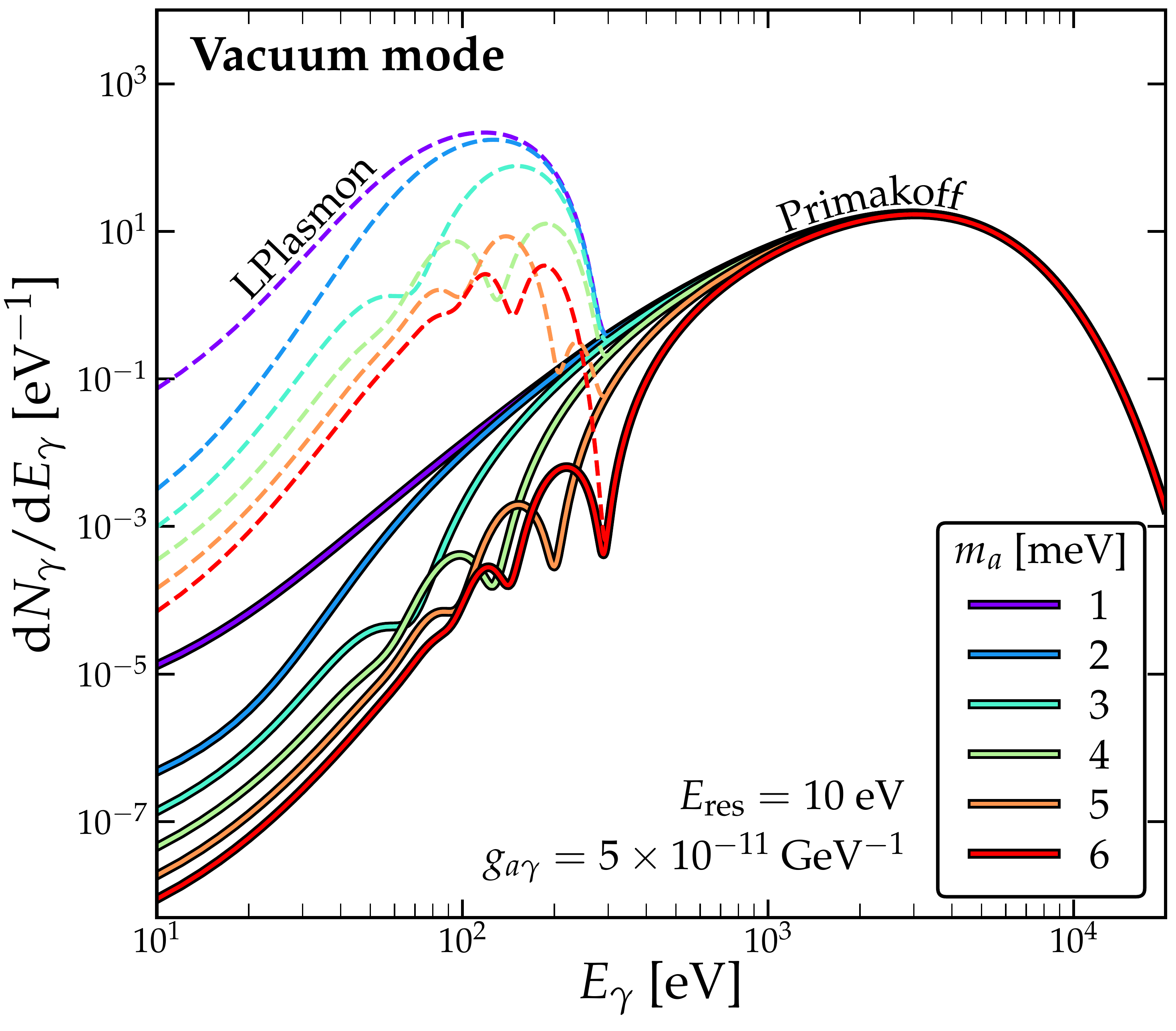}
\includegraphics[width=0.49\textwidth] {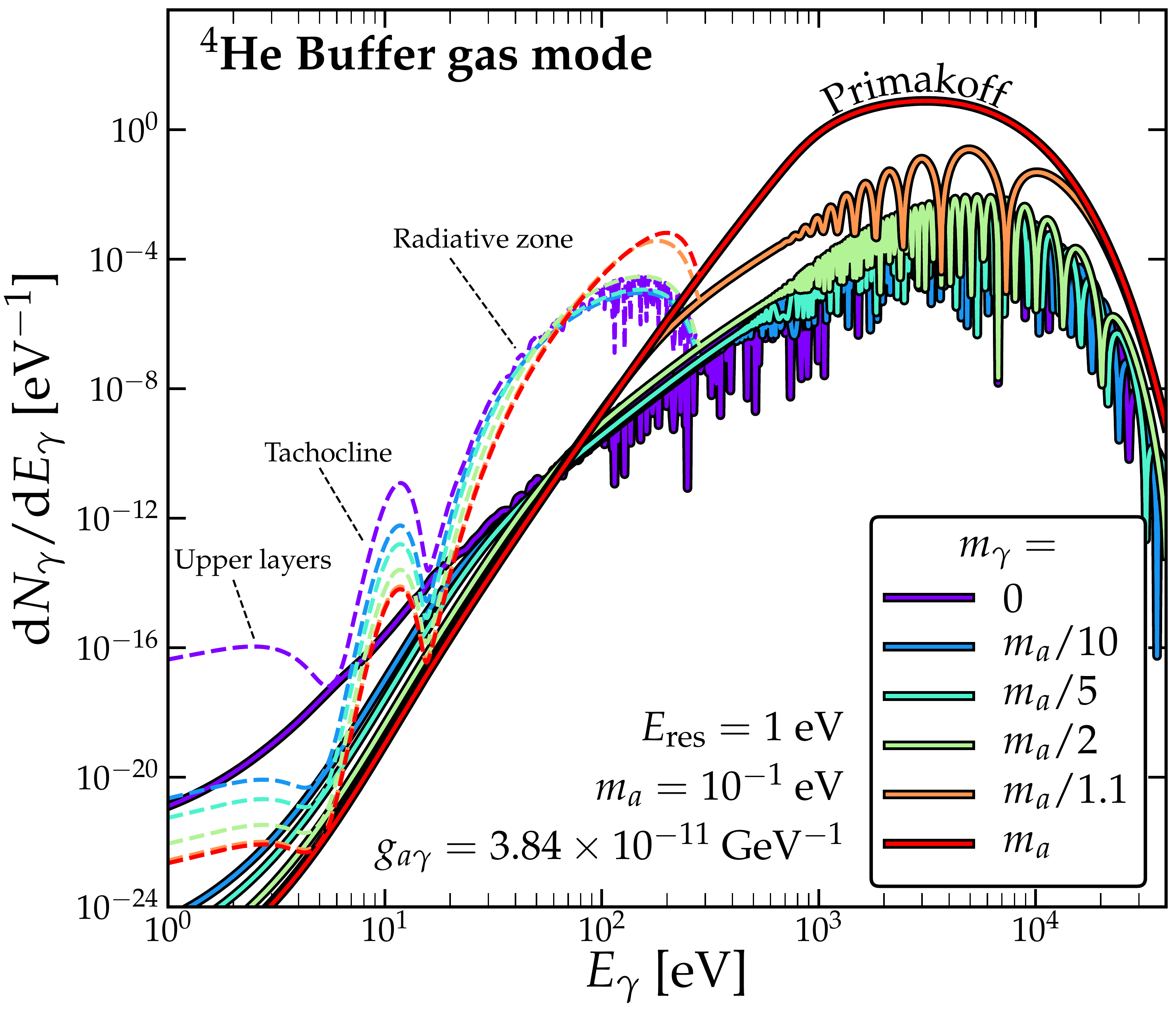}
\caption{\label{fig:XraySpectrum_lowmasses} Photon spectra in IAXO over a range of axion masses in the vacuum mode (left) and for a single mass but a range of pressures in the buffer gas mode (right). The solid lines show the contribution to the signal from the Primakoff flux alone, and the dashed lines indicate the upper bound of the contribution from the LPlasmon flux assuming the seismic solar model. For the left-hand panel we have selected a particular small range of masses for which the additional flux raises the small differences between the spectra. Over this range, the LPlasmon flux provides an additional discriminant on the value of $m_a$; which would require a much higher value of $\gag$ to measure otherwise. On the right-hand panel we express the pressure settings in terms of $m_\gamma/m_a$. The indigo line, $m_\gamma = 0$, is equivalent to a vacuum, and the red line, $m_\gamma=m_a$, is equivalent to $p_{\rm max}$: the pressure that maximizes the overall signal. Notably, the lowest energy part of the spectrum is amplified rather than suppressed as the pressure is decreased, we discuss this subtlety further in the text. On the left we assume a realistic energy resolution of 10 eV, and on the right we assume a much more optimistic 1 eV which allows us to display the spectra from the solar $B$-field all the way up to the upper layers.}
\end{figure*}
In the left-hand panel of Fig.~\ref{fig:XraySpectrum_lowmasses} we show the resulting photon spectra for both the Primakoff and LPlasmon fluxes. We assume an energy resolution of 10 eV, so only the LPlasmon flux from the radiative zone is visible. In fact for this mass range the additional number of events due to this flux can be quite large: between $24800$--$300$ for $m_a~=~1$--$6$~meV and $\gag = 5\times 10^{-11}$~GeV$^{-1}$. Notice also that the different spectra for different masses diverge from each other only at energies below $\sim 500$~eV, precisely where the LPlasmon flux is amplifying the signal. This observation will be relevant later when we discuss how the LPlasmon flux can enhance IAXO's ability to measure the value of $m_a$ in its vacuum mode.

On the right-hand panel of Fig.~\ref{fig:XraySpectrum_lowmasses} we show a set of spectra for the buffer gas mode. Here we fix the axion mass at $m_a = 10^{-1}$~eV and instead vary $m_\gamma$ between 0 and $m_a$, effectively varying the pressure between a vacuum and $p_{\rm max}$: the maximum pressure required to restore the axion-photon dispersion relations. When $m_a = m_\gamma$ the total number of photons is maximized, and with decreasing pressures (lower $m_
\gamma$) the spectrum is suppressed. However not all of the spectrum is suppressed evenly by the axion-photon interference, in fact the lowest energy components are enhanced as the pressure is lowered down to a zero.

The origin of this peculiarity can be understood by considering the energy dependence of $P_{a\rightarrow\gamma}(\omega)$ in Eq.\eqref{eq:Pbuffergas}. For a given value of $m_a$ and a pressure somewhere in between zero and $p_{\rm max}$, the probability will scale with $\omega^2$ at very low energies, and then as $\omega^{6.2}$ at higher energies (because of the energy dependence of $\Gamma$). The regimes cross over when $q^2 \sim \Gamma^2/4$, at an energy $\omega_{\rm cross} \propto p^{1/2.1}$. So for accessing features at very low energies, it is more favorable to have $\omega_{\rm cross}$ be lower so that the desired part of the spectrum is in the regime $\omega>\omega_{\rm cross}$ where $P\sim \omega^2$. For the sub-100 eV spectrum, the vacuum setting is optimal, even for large axion masses. This is why the dashed lines on the right-hand of Fig.~\ref{fig:XraySpectrum_lowmasses} continue to rise at low energies as the pressure is decreased, even as the high energy part of the spectrum becomes suppressed by oscillations. The oscillations begin at $\omega_{\rm osc}$ which corresponds to when $\exp{(-\Gamma L)} \sim 2\exp{(-\Gamma L/2)}$ in Eq.\eqref{eq:Pbuffergas}. Over all axion masses $\omega_{\rm osc}>\omega_{\rm cross}$.

These arguments inform to us that even though the buffer gas mode is essential for discovering the higher mass axions, a low pressure or vacuum mode is still desirable for studying very low energy components like the upper layers or the tachocline. The radiative zone is in an intermediate energy range and the pressure must be chosen so that $\omega_{\rm cross}\lesssim 20$~eV and $\omega_{\rm osc}\gtrsim 300$~eV.

\subsection{Background}
The background level in the IAXO detectors is expected to be extremely low at around $10^{-8}$ to $10^{-9}$~cm$^{-2}$~s$^{-1}$~keV$^{-1}$ \cite{Irastorza:2013dav, Armengaud:2014gea}, amounting to only a few counts in the signal region of interest over the full data-taking campaign. The number of events due to the background is given by,
\begin{equation}\label{eq:background}
    N_{\rm bg} = A_{\rm det} t (E_{\rm max}-E_{\rm res}) \Phi_{\rm bg}
\end{equation}
where the total detector area is $A_{\rm det}$ and $\Phi_{\rm bg}$ is the estimated background flux (both given in Table~\ref{tab:IAXOparams}). We assume that the background does not depend on energy. This assumption is justified from an experimental standpoint~\cite{Armengaud:2014gea}; it is also unlikely that our results will be sensitive to it either, since there are no background sources that would mimic an axion signal.

\subsection{Search strategy}\label{sec:strategy}
The precise search strategy envisaged for IAXO has not yet been finalized. In principle there are many potential options which could target certain mass ranges. To have some concreteness, we have simply implemented a search strategy which attempts to replicate the most recent published projections made by the IAXO Collaboration~\cite{Armengaud:2019uso}. We assume that a 6 year total data-taking time is split evenly between three phases:
\begin{itemize}
    \item {\bf Vacuum phase}: fixed vacuum pressure mode with no scanning, sensitive to the full range of axion masses, but with a loss in sensitivity above $\sim 10^{-2}$~eV.
    \item {\bf Intermediate buffer gas phase}: scanning over linearly spaced pressures between $p=0.036$ and $0.44$~mbar. This gives approximately horizontal sensitivity to $\gag$ over a range of axion masses that bridges the gap between the KSVZ line at $2\times 10^{-2}$~eV and the DFSZ~II line at $7\times 10^{-2}$~eV.
    \item {\bf High mass buffer gas phase}: pressure values between $p = 0.44$ and $2.916$~mbar that are spaced proportionally to $1/p$. This maintains DFSZ II sensitivity between $7\times 10^{-2}$ and $1.8\times 10^{-1}$~eV.
\end{itemize}
Helium-4 begins to condense above 16.405 mbar at $T_{\rm gas} = 1.8$~K, corresponding to maximum mass for the buffer gas phase of $m_a \approx 0.4$~eV. Fortunately QCD axions above this mass are already excluded by a stellar evolution bound using horizontal branch stars in globular clusters~\cite{Ayala:2014pea}. So we do not need to worry about switching the buffer gases to $^3$He, as was the case with CAST~\cite{Arik:2011rx}.
 
\begin{figure}[t]
\includegraphics[width=0.49\textwidth] {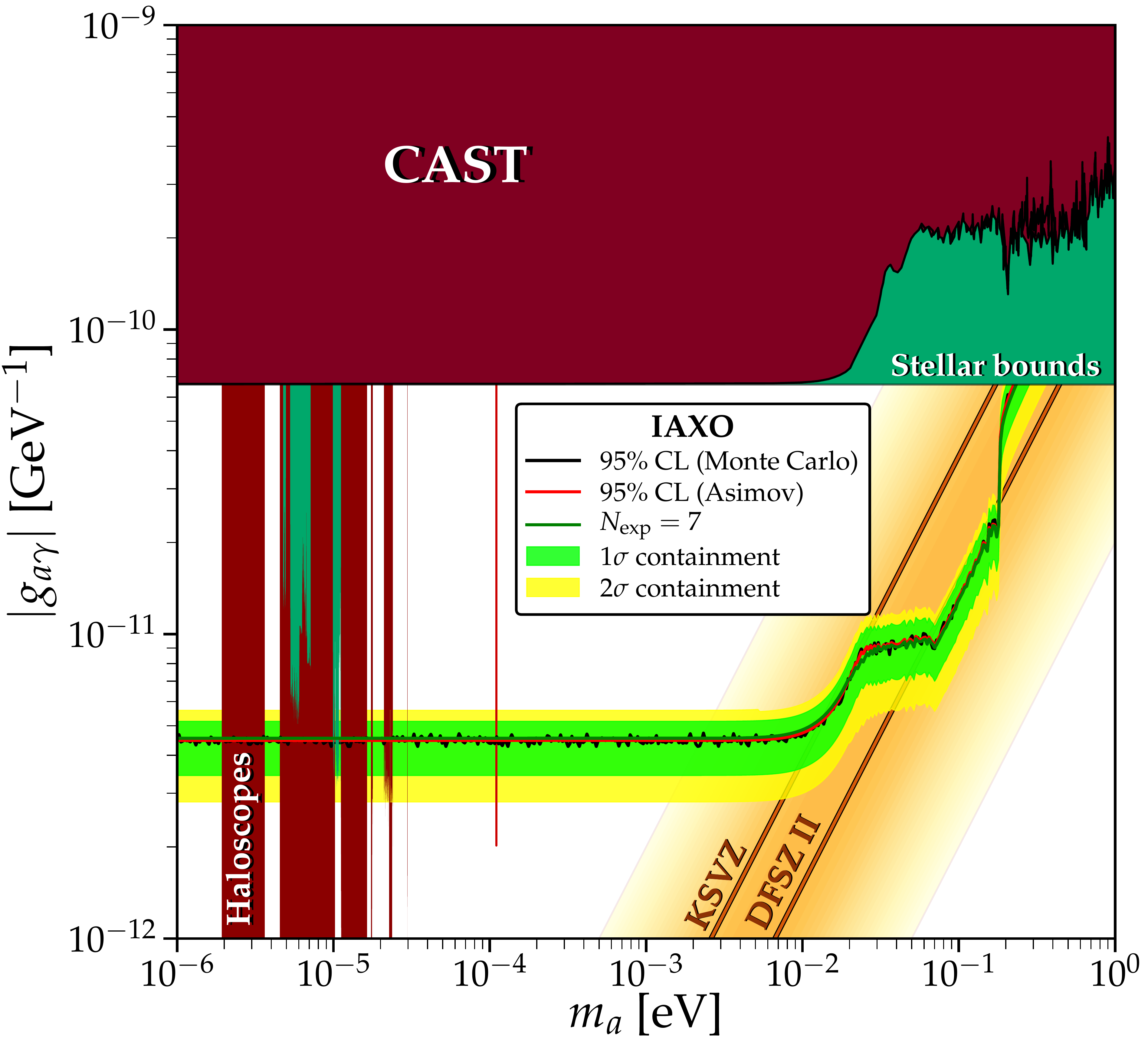}
\caption{\label{fig:IAXO_Sensitivity_Asimov} Expected 95\% CL exclusion limit projected for IAXO, as calculated via our profile likelihood ratio test. We show the result from a toy Monte Carlo simulation of the test on mock data (black line) as well as the result using Asimov data (red). The projection is roughly equivalent to a constant expected event number line for $N_{\rm exp} = 7$ (shown as a green line). In this plot and subsequent ones we also show the relevant existing constraints on $\gag$. Experimental constraints are shown in dark red and include the results from CAST~\cite{Anastassopoulos:2017ftl} and dark matter axion haloscopes~\cite{Asztalos:2010,DePanfilis:1987dk, Hagmann:1990tj, Zhong:2018rsr,Du:2018uak,McAllister:2017lkb,Alesini:2019ajt,Braine:2019fqb}. Astrophysical constraints are shown in green: the stellar bound from horizontal branch stars in globular clusters~\cite{Ayala:2014pea}, and a recent indirect dark matter search with radio observations of neutron stars~\cite{Foster:2020pgt}.}
\end{figure}

For our analysis we will assume that the data from the three phases are combined and analyzed jointly. The respective sensitivities of each are thereby allowed to overlap. Our reproduction of IAXO's sensitivity is shown in Fig.~\ref{fig:IAXO_Sensitivity_Asimov}. The statistical approach we have used to derive this limit is explained in Sec.~\ref{sec:stats}.

\section{Results}\label{sec:results}

\subsection{Axion mass sensitivity}
\begin{figure}[t]
\includegraphics[width=0.49\textwidth] {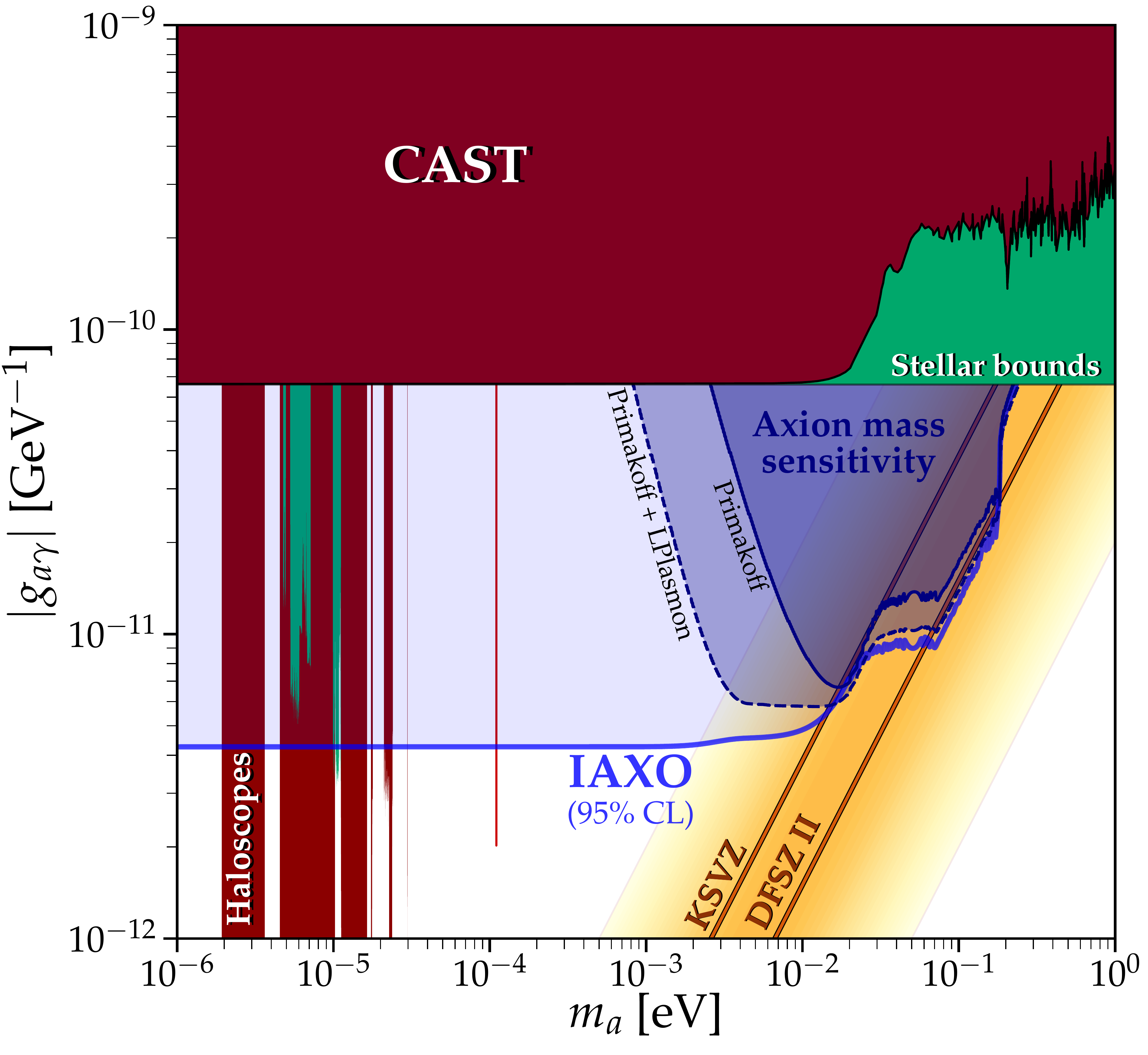}
\caption{\label{fig:IAXO_mass_sensitivity} IAXO's expected sensitivity to a 3$\sigma$ acceptance of a nonzero axion mass for two different assumptions for the incoming axion flux. The signal becomes increasingly insensitive to the value of $m_a$ below $10^{-2}$~eV, but when the low energy LPlasmon flux is included (dashed line), the additional events at low energies means the mass can be discriminated from zero down to lower values. We assume a benchmark magnetic field which lies in the middle of our window of models, Eq.\eqref{eq:Bfield_window}. We also show for comparison IAXO's projected 95\% CL sensitivity to $\gag$ from Fig.~\ref{fig:IAXO_Sensitivity_Asimov}.}
\end{figure}

\begin{figure*}[t]
\includegraphics[width=0.49\textwidth] {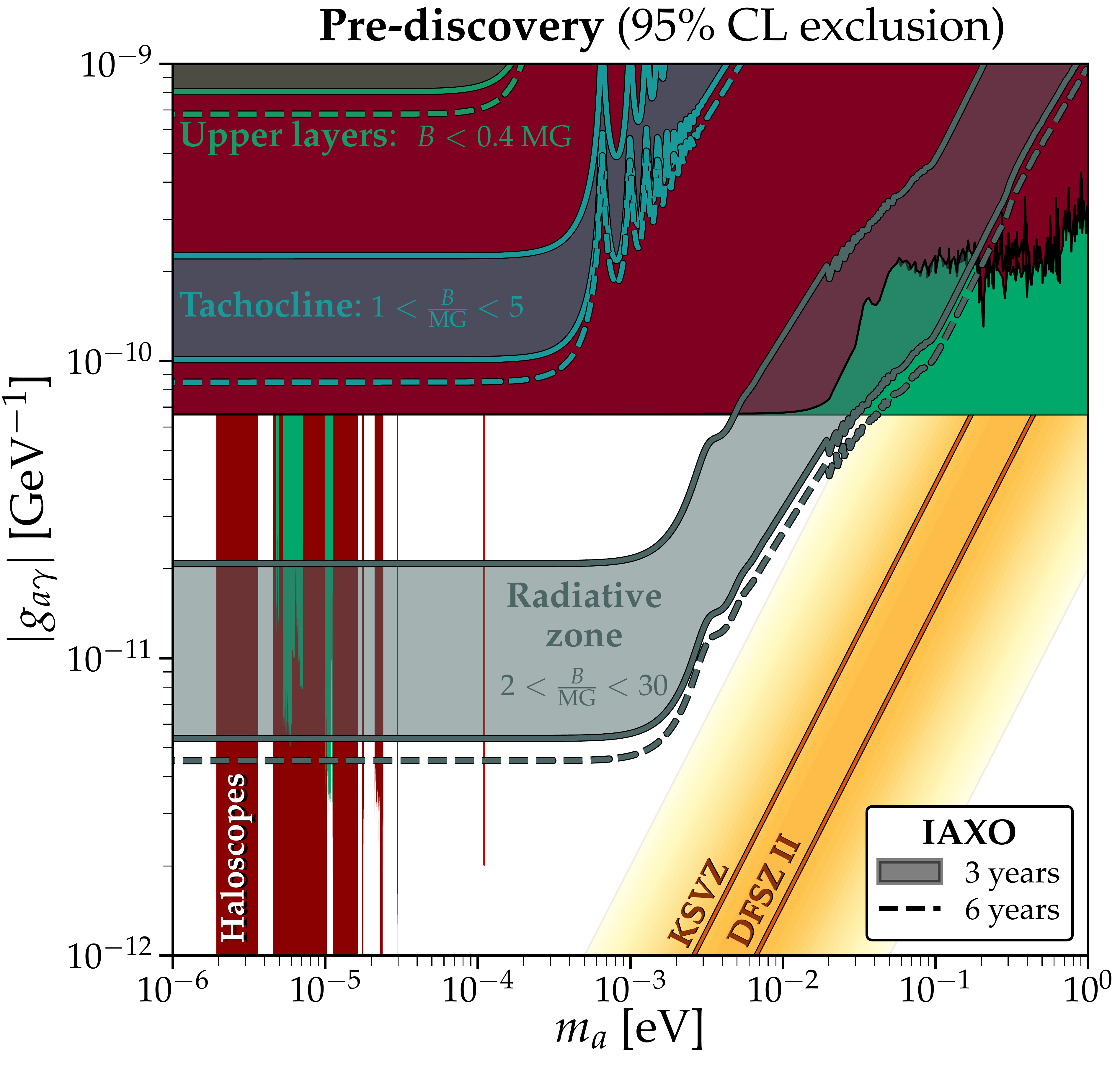}
\includegraphics[width=0.49\textwidth] {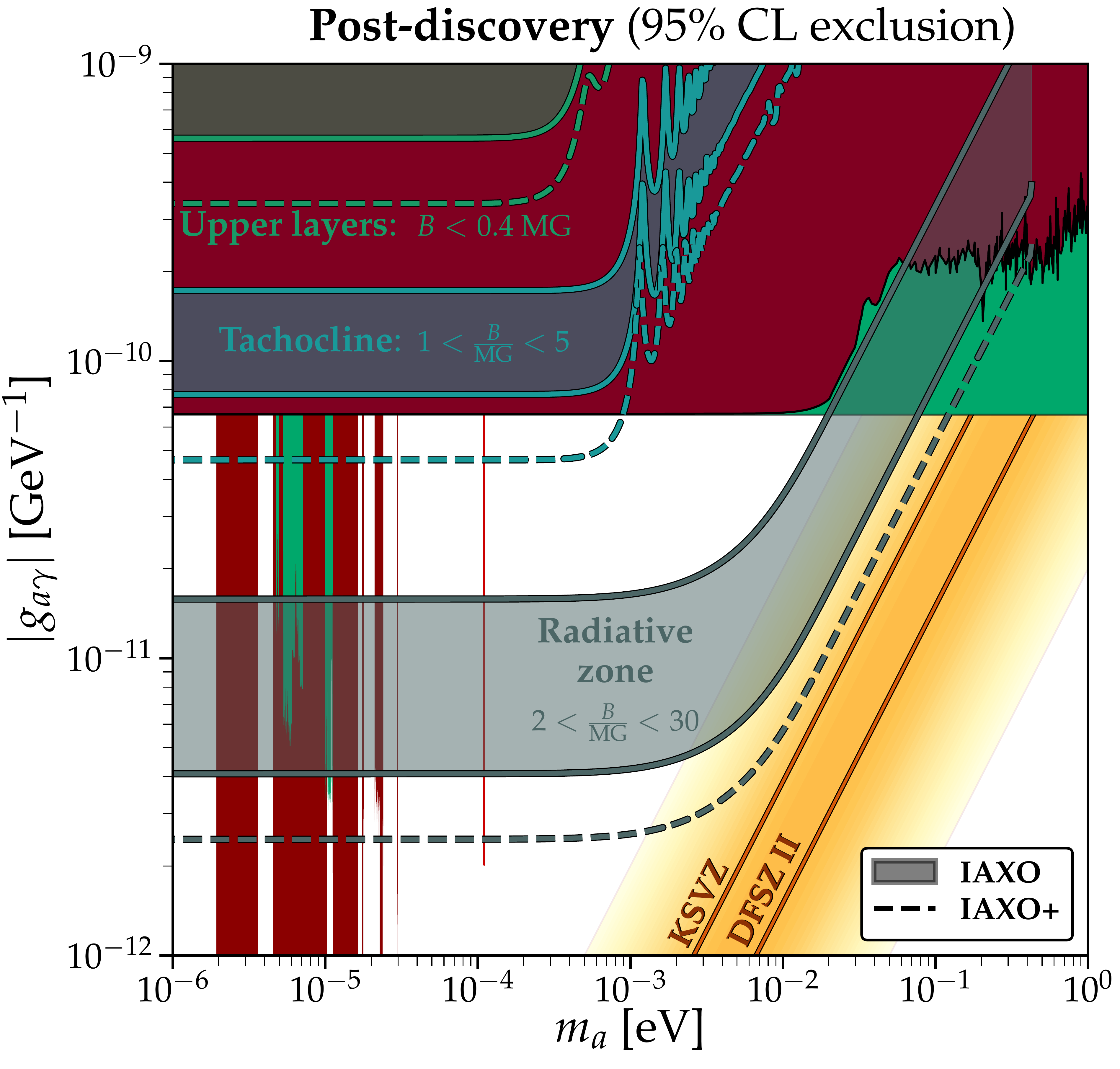}
\caption{\label{fig:IAXO_Bfield_sensitivity} Median sensitivity at 95\% CL to a nonzero value of the magnetic field in the three main regions we study here: the radiative zone, tachocline and the upper layers. The shaded regions show the sensitivities over the range of field strengths in Eq.\eqref{eq:Bfield_window}. The left and right-hand panels correspond to two possible data-taking campaigns. The left-hand panel corresponds to the ``pre-discovery sensitivity'', which is the $B$-fields that could be constrained with an exposure the same duration as the planned axion search with IAXO (with an extended 6 year duration indicated with a dashed line). On the other hand the right-hand panel shows the ``post-discovery sensitivity'' which corresponds to a scenario in which the axion is detected and then IAXO continuously observes the solar axion flux for 6 years with the optimum pressure setting. The dashed line in this case corresponds to the improvement in the lower edge of the shaded regions when we assume the experimental parameters of IAXO+.}
\end{figure*}

\begin{figure*}[t]
\includegraphics[width=\textwidth]{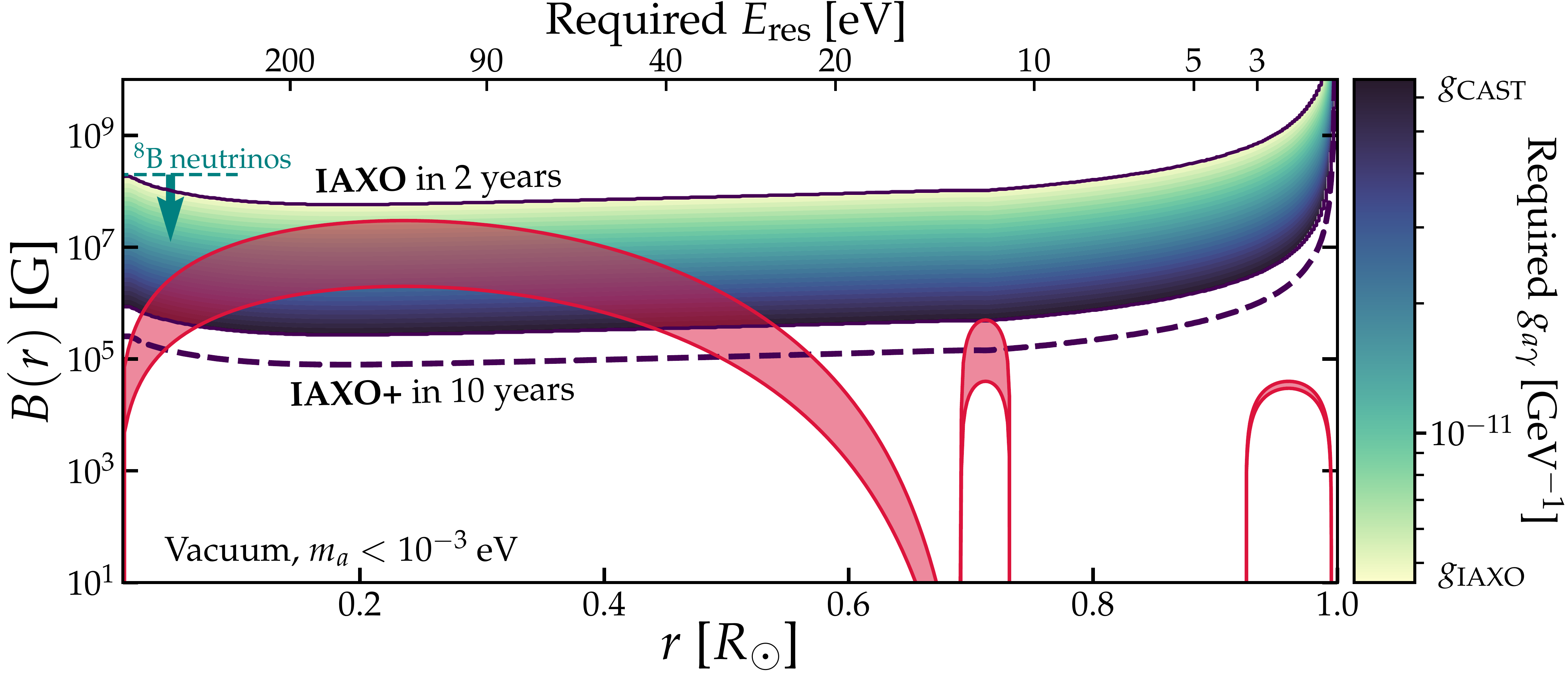}
\caption{\label{fig:Bfield_sensitivity} Sensitivity of IAXO to the solar magnetic field profile. The red bands indicate our range of magnetic field models described in Sec.~\ref{sec:solarBfield}, whereas the blue-green colored region shows the minimum value of $\gag$ required to reach a particular value of $B(r)$. We bound the region between a maximum value given by the value of $\gag$ already excluded by CAST: $g_{\rm CAST} = 6.6\times 10^{-11}$~GeV$^{-1}$ (i.e. IAXO cannot probe the field if it requires a coupling larger than what is already excluded) and a minimum value given by the projected sensitivity of a 2 year IAXO exposure: $g_{\rm IAXO} = 4.5\times10^{-12}$~GeV$^{-1}$ (i.e.~IAXO cannot probe the field for couplings smaller than its sensitivity). So for example, in 2 years IAXO can probe our range of models for radii $r\lesssim 0.5 R_\odot$. As a dashed line we show the improvement in sensitivity if we assume a 10 year exposure with the upgraded configuration IAXO+. Since accessing the field at larger radii requires sensitivity to smaller photon energies, we also show on the upper horizontal axis the maximum required energy resolution as function of $r$.}
\end{figure*}

One of the challenges faced by helioscope experiments is the measurement of the axion mass. Axion masses in the buffer gas regime are straightforwardly measurable because the signal is maximized when the effective photon mass (given by the gas pressure) matches the axion mass. On the other hand, axions lighter than an meV are essentially indistinguishable from being massless. So the axion mass is not a determinable parameter for much of the parameter space. However there is an intermediate regime as discussed in Ref.~\cite{Dafni:2018tvj} where the oscillatory features that are present in the signal for $m_a\gtrsim$~meV could be used as a handle on the axion mass. This allows IAXO to extend its sensitivity to a nonzero $m_a$ down to around $10^{-2}$~eV. For masses below this, the part of the photon spectrum that contains information about the axion mass gets rapidly pushed below any semi-realistic energy threshold. In light of the low energy LPlasmon flux however we now have an additional source of signal at very low energies. This means that the axion mass discrimination should be possible to lower masses than when only the Primakoff flux is accounted for.

In Fig.~\ref{fig:IAXO_mass_sensitivity} we show IAXO's sensitivity to a nonzero value of the axion mass. We do this by calculating expected discovery limits for a 3$\sigma$ acceptance of $m_a>0$ in 50\% of experiments. We compare the mass discovery limit under the assumption of solely the Primakoff flux (solid line), with the same limit assuming both the Primakoff and LPlasmon flux from the seismic solar model. We see that indeed the axion mass sensitivity can be pushed down to masses almost an order of magnitude lower. The axion mass sensitivity is also improved in the buffer gas mode. Although the LPlasmon flux is a negligible increase in signal for these masses, because we conduct the test against the massless hypothesis which does have a sizeable signal from the LPlasmon flux, the mass discrimination requires fewer overall events.

This increase in sensitivity assumes that the LPlasmon flux is accounted for. Therefore this result is model dependent. Nonetheless, the distinction between the limits for the two different flux assumptions is still relevant: it tells us that the massless axion hypothesis looks significantly different to the massive case when the LPlasmon flux is present. The LPlasmon flux is certainly going to be present since the Sun must have a magnetic field. So whichever model is assumed, the mass discovery will always require fewer overall events than has been considered in the past.


\subsection{B-field sensitivity}
In Fig.~\ref{fig:IAXO_Bfield_sensitivity} we show the sensitivity of IAXO to the three regions of the solar magnetic field profile individually. For completeness we assume an energy resolution of 2 eV here so that we can show results for the magnetic field of the upper layers. The sensitivity to the radiative zone and the tachocline are generally unaffected by this choice as long as we use values $E_{\rm res}\lesssim30$ and $10$~eV respectively.

The sensitivities plotted here are the projected median 95\% CL exclusion for a nonzero value of the magnetic field, where each set of curves corresponds to testing for $B_{\rm rad}$, $B_{\rm tach}$ or $B_{\rm upper}$ individually. Since these regions do not overlap in energy, when testing for one field normalization, we can simply ignore the spectrum from the other two. 

A particular sensitivity line on these plots corresponds to a particular value of magnetic field strength normalization and should be interpreted in the following way: if the assumed magnetic field strength used to make the limit is say, $B_1$, then IAXO is can constrain all values of $B>B_1$ at 95\% CL for all axion masses and couplings lying \emph{above} the sensitivity line drawn.  The shaded regions correspond to the same range of assumed field normalizations as shown in previous figures.

To fully demonstrate the full potential of IAXO we define two different strategies for the left and right-hand panels of Fig.~\ref{fig:IAXO_Bfield_sensitivity}. The left-hand, ``pre-discovery'' scenario corresponds to the $B$-field sensitivity of IAXO with a 3 year (6 year for the dashed line) data-taking exposure following the IAXO search strategy outlined previously in Sec.~\ref{sec:strategy}, and used in both Fig.~\ref{fig:IAXO_Sensitivity_Asimov} and~\ref{fig:IAXO_mass_sensitivity}. That is to say: if IAXO detects the axion in its initial search campaign, it will already have the sensitivity to the solar magnetic profile as shown in the left-hand panel. The tachocline is therefore not observable in the pre-discovery scenario, but for a large range of axions with $\gag\gtrsim 6\times 10^{-12}$~GeV$^{-1}$ constraints could be placed already on the radiative zone's magnetic field.

Measuring the solar magnetic field is truly part of the post-discovery physics case for IAXO. So in the right-hand panel we show enhanced sensitivities which assume a 6 year data-taking exposure with either IAXO or IAXO+, in which the buffer gas pressure is fixed to the optimum value required to measure each component of the $B$-field. For the radiative zone this optimum value is the pressure for which $m_a=m_\gamma$, whereas for the tachocline and upper layers a vacuum is the optimum pressure over all masses.

The radiative zone is easily within IAXO's reach, and could even be constrained well into the QCD axion model band just above KSVZ. The tachocline is mostly out of reach unless $\gag$ is only just below CAST. Such an axion would be detected very soon into IAXO's initial campaign (and probably by babyIAXO), but it would still take 6 years with an upgraded magnet to begin to constrain the tachocline. We emphasize that we have only fixed a benchmark data-taking exposure of 6 years here, the sensitivity could continue to even smaller couplings for a longer duration, scaling as $\gag \sim t^{-1/4}$. This kind of measurement therefore provides considerable motivation for continuing IAXO beyond its initial discovery of the axion.

\subsection{Sensitivity versus radius}
Finally, to summarize the sensitivity to the magnetic field in a simple, digestible way we present Fig.~\ref{fig:Bfield_sensitivity}, which shows again the range of magnetic field profiles in detail, along with IAXO's sensitivity as a function of radius and $E_{\rm res}$.  This plot demonstrates that even for modest energy resolutions, IAXO is readily sensitive to a wide range of well-motivated magnetic field strengths in the inner 50\% of the Sun. Probing up to 0.5~$R_\odot$ would require energy resolutions better than $\sim$30 eV, which is already considerably larger than the demonstrated $\sim 2$~eV resolutions of metallic magnetic calorimeters for instance~\cite{Kempf:2018}. Interestingly, we also find that solar axions could easily constrain the magnetic field of the solar core to smaller values than is currently possible with the temperature dependence of the $^8$B neutrino flux. We also see that for axion couplings on the larger side, IAXO+ would even be able to begin to probe the tachocline. The sensitivity shown in Fig.~\ref{fig:Bfield_sensitivity} applies consistently to all axion masses below approximately $10^{-3}$~eV.

\section{Conclusions}\label{sec:conc}
Being the only example of a star close by, the Sun provides our only prototypical example of a stellar magnetic field. Therefore it is the basis upon which must try to understand the role of magnetic fields in the formation, structure and evolution of cool stars across the Universe~\cite{Solanki_review}. Yet the structure of the solar B-field remains poorly understood~\cite{Friedland:2002is,Antia2000,2002ApJ...578L.157C,Baldner_2008,Baldner_2009,Kiefer_2018,Antia2000,Baldner_2009,Baldner_2008,2002ApJ...578L.157C,Duez:2009vg,Maeder:2003ym}. Most existing measurements of the field are limited over certain regions and often rely on a variety of modeling assumptions.
In this paper we have aimed to determine the potential for axion helioscopes to serve a dual purpose as instruments to measure the magnetic field of the Sun. While there are already several ways to constrain the Sun's magnetic field --- including both solar physical measurements such as helioseismology, and astroparticle probes like solar neutrinos --- there is no single method to map the entirety of the field profile. 

Axions would therefore be a greatly welcomed new handle on the solar magnetic field. Axions converting from longitudinal plasmons at a given position inside the Sun have an energy equal to the plasma frequency at that position, and a flux proportional to the square of the magnetic field. In other words, the number of signal photons from axions converting back into photons inside a helioscope provides information about the strength of the magnetic field, and the energies of those photons provide a measure of the positions they originated from. Therefore a measurement of this flux of axions could be almost directly recast as a measurement of the magnetic field profile.

We have centered our experimental configuration around the baseline and upgraded versions of IAXO as summarized in Table~\ref{tab:IAXOparams}. However the main uncertainty regarding the final specification of IAXO is the energy resolution of its detectors. Measuring the magnetic field at larger solar radii requires smaller energy resolutions. So we have framed our discussion around the resolution required to probe out to a certain solar shell. This main result is summarized in Fig.~\ref{fig:Bfield_sensitivity}. If we assume the most optimistic energy resolution of say metallic magnetic calorimeters $\sim$~2 eV, we could expect IAXO or IAXO+ to be sensitive to the inner 70\% of the Sun and potentially measure the field out to the tachocline. For axions heavier than $\sim 10^{-3}$~eV the sensitivity steeply declines as shown in the left-hand panel of Fig.~\ref{fig:IAXO_Bfield_sensitivity}. In a post-discovery campaign in which the pressure setting could be optimized to focus on certain parts of the axion spectrum, we find that IAXO+ would be able to constrain the magnetic field of the radiative zone even for couplings within the QCD axion band. This is shown in the right-hand panel of Fig.~\ref{fig:IAXO_Bfield_sensitivity}.

An interesting side effect of the longitudinal plasmon flux is that it can improve IAXO's ability to measure the value of the axion mass, as shown in Fig.~\ref{fig:IAXO_mass_sensitivity}. This result assumes that the flux from the radiative zone is correctly modelled in our likelihood analysis, so the enhanced sensitivity is model dependent. Nevertheless, the LPlasmon flux is guaranteed at some level as long as the Sun has a magnetic field, which it demonstrably does. Therefore the extra source of axions we have studied here is not only beneficial with regards to measuring the magnetic field but would also help us in determining the axion's properties. Moreover, this could go beyond simply the mass: since the LPlasmon flux is only dependent on $\gag$, it could also help in distinguishing the axion-photon and axion-electron fluxes for non-hadronic QCD axion models (see e.g.~Ref.~\cite{Jaeckel:2018mbn}), though we have not explored that possibility here.

IAXO may also have sensitivity to the angular structure of the solar magnetic field, rather than just the radial profile. In Fig.~\ref{fig:Modulation} we showed that there is a 1.6\% biannual modulation present in the flux from a toroidal magnetic field. We have calculated the time dependence of this modulation over the year so that it can be of future use. A more detailed exploration of how such a modulation signal could be used to constrain the angular profile may reveal more of IAXO's future potential as a solar magnetometer. This would probably require many more events to make use of than our benchmark cases, and perhaps a helioscope even larger than IAXO. Though in the event of a detection of solar axions this expense might be much more easily spared.

\acknowledgments
We thank Sarbani Basu and Aldo Serenelli for providing us with additional literature on the solar magnetic field, and Javier Redondo for helpful correspondence. CAJO is supported by the University of Sydney and the Australian Research Council. AC acknowledges support from the “Generalitat Valenciana” (Spain) through
the “plan GenT” program (CIDEGENT/2018/019), as
well as national grants FPA2014-57816-P, FPA2017-
85985-P. AM is supported by the
European Research Council under Grant~No.~742104 and is supported in part by the research environment grant ``Detecting Axion Dark Matter In The Sky And In The Lab (AxionDM)" funded by the Swedish Research Council (VR) under Dnr 2019-02337. The  work  of EV  was  supported  by the U.S. Department of Energy (DOE) Grant No.  DE-SC0009937.

\appendix

\section{Statistical methodology}\label{sec:stats}
Our statistical methodology which is used to derive all limits in this paper is based on a frequentist profile likelihood ratio test between a null and alternative hypothesis, $H_{0,1}$. In order to use asymptotic limits for likelihood ratios, we will assume that $H_0$ is nested within a more general $H_1$. Furthermore, to facilitate the later use of Asimov data, we adopt a binned likelihood for the model, $\mathscr{L}$ which is the product of Poisson probability distribution functions $\mathscr{P}$ for $N_{\rm obs}$ photons, given an expected number $N_{\rm exp}$, for each of the $N_{\rm bins}$ bins:
\begin{equation}\label{eq:Like}
\mathscr{L} = \prod_{i=1}^{ N_{\rm bins}}\mathscr{P}\left[N_\textrm{obs}^i \bigg| N^i_{\rm exp}(m_a,\,\gag,\, \mathcal{B}) + N^i_{\rm bg}(\Phi_{\rm bg})\right] \, .
\end{equation}
The expected number of photons $N_{\rm exp}$ in bin $i$ is obtained by (1) calculating the full spectrum using Eq.\eqref{eq:Ngamma}; (2) applying the Gaussian energy resolution kernel with width $E_{\rm res}$; and then (3) integrating the convolved spectrum over the $i$-th bin. Throughout we assume that the background level is flat and that $N^i_{\rm bg}$ is given by Eq.\eqref{eq:background} but with the energy interval replaced by the bin width.

First, to calculate IAXO's projected exclusion limit on $\gag$, we define a test statistic using the likelihood ratio --- at fixed values of $m_a$ and 
magnetic field parameters $\mathcal{B}$ --- which tests for $H_1$ when $\gag>0$ (signal+background) against $H_0$ which has $\gag=0$ (background only),
\begin{equation*}\label{eq:llhoodratio}
    q(m_a) = -2\ln \left[\frac{ \mathscr{L} (m_a,0,\mathcal{B},\hat{\hat{\Phi}}_{\rm bg}) }{\mathscr{L} (m_a,\hat{g}_{a\gamma},\mathcal{B},\hat{\Phi}_{\rm bg})} \right] \, .
\end{equation*}
Where the $\hat{}$ parameters are the maximum likelihood estimators (MLEs) under $H_1$ and the $\hat{\hat{}}$~parameters are the MLEs under $H_0$. Parameters without hats are fixed. The asymptotic distribution of $q$ under $H_0$ is a ``half-chi-squared'' distribution: $\frac{1}{2}\delta(0) + \frac{1}{2}\chi^2_{1}$, according to Chernoff's theorem~\cite{Chernoff:1954eli} --- a special case of Wilks' theorem~\cite{Wilks:1938dza} that is applied when testing for a parameter at the boundary of its allowed space (see e.g.~Ref.~\cite{Algeri:2019arh} for more discussion of this distinction). To calculate the expected 95\% CL exclusion limit that could be set on $\gag$ by IAXO, we perform Monte Carlo simulations using mock datasets to obtain the distributions of $q$ under $H_{0,1}$.

While doing this, we can also validate an asymptotic limit of the test using Asimov data, which might free us of performing more Monte Carlo simulations. Asimov data is the name given to a set of mock data which exactly matches the expectation, i.e.~$N^i_{\rm obs} = N^i_{\rm exp}+N^i_{\rm bg}$ over all bins. When performing the likelihood ratio test on the Asimov dataset, the resulting value of $q$ quickly approaches the median of the true distribution as the number of (signal or background) events increases~\cite{Cowan:2010js}. Although this approximation only holds exactly in the limit of high statistics, a reasonable estimate of the median $q$ is reached even for relatively low numbers of events $\mathcal{O}(10)$. In order to infer the significance of the Asimov result without generating the correct distribution of $q$ under $H_0$ we can use another asymptotic limit: the significance of a particular test result $q^*$ under $H_1$ is approximately $\sqrt{q^*}$. Since the null distribution is a half-chi-squared, a confidence level of 95\% ($p$ value of 0.05) corresponds to $\sqrt{q^*} = 1.64$.

In Fig.~\ref{fig:IAXO_Sensitivity_Asimov} we show our estimation of the expected one-sided 95\% CL exclusion limit that could be set by IAXO. With a black line we show the result from the full Monte Carlo simulation of the null and alternative distributions of $q$, assuming no asymptotic results. Then as a red line we show the Asimov approximation which obtains extremely good agreement. We also show the expected $1$ and $2\sigma$ containment of the median 95\% CL exclusion limit which can also be obtained straightforwardly with the Asimov dataset. For reference we also show in green a contour of values of ($\gag,m_a$) for which the expected number of events is $N_{\rm exp} = 7$. Since the spectral shape of the axion signal does not change much with mass, the exclusion limit is roughly (though not precisely) the same as a constant event number contour. 
All of our remaining results in the next section are obtained in a very similar way. For example we can find the projected limits for the axion mass by floating $m_a$ in Eq.\eqref{eq:llhoodratio} and testing for the hypothesis of $m_a>0$ against the case $m_a = 0$. The validity of Chernoff's theorem and the Asimov dataset for this exact scenario were already validated in Ref.~\cite{Dafni:2018tvj}. 

We can also find projected limits for the exclusion of nonzero values of any parameter in $\mathcal{B}$ describing the $B$-field in the same way. To have the $\mathcal{O}(>10)$ events needed to measure the LPlasmon flux, one will necessarily also have many more events from the Primakoff flux. So it is safe to assume that the Chernoff and Asimov assumptions will be valid here too.

 
\bibliographystyle{bibi.bst}
\bibliography{IAXOPlasmon}

\end{document}